\pdfoutput=1

\documentclass[11pt,twoside,a4paper,cmspaper,final,collab]{cms-tdr}
\def\svnVersion{456200P}\def\cmsCernNoTag{CERN-EP-2017-257}\def\cmsCernDate{\today}\def\cmsMessage{Published in the European Physical Journal C as \href{http://dx.doi.org/10.1140/epjc/s10052-018-5752-x}{\doi{10.1140/epjc/s10052-018-5752-x.}}}

\begin{document}\cmsNoteHeader{SMP-15-009}

\hyphenation{had-ron-i-za-tion}
\hyphenation{cal-or-i-me-ter}
\hyphenation{de-vices}

\RCS$Revision: 448929 $
\RCS$HeadURL: svn+ssh://svn.cern.ch/reps/tdr2/papers/SMP-15-009/trunk/SMP-15-009.tex $
\RCS$Id: SMP-15-009.tex 448929 2018-03-02 21:39:14Z alverson $

\newlength\cmsFigWidth
\ifthenelse{\boolean{cms@external}}{\setlength\cmsFigWidth{0.85\columnwidth}}{\setlength\cmsFigWidth{0.4\textwidth}}
\newlength\cmsSingleFigWidth
\ifthenelse{\boolean{cms@external}}{\setlength\cmsSingleFigWidth{0.40\textwidth}}{\setlength\cmsSingleFigWidth{0.45\textwidth}}
\newlength\cmsDoubleFigWidth
\ifthenelse{\boolean{cms@external}}{\setlength\cmsDoubleFigWidth{0.35\textwidth}}{\setlength\cmsDoubleFigWidth{0.45\textwidth}}
\newcommand{\x}{\ensuremath{\phantom{0}}}
\newcommand{\y}{\ensuremath{\phantom{.}}}
\providecommand{\PX}{\ensuremath{X}\xspace}

\ifthenelse{\boolean{cms@external}}{\providecommand{\cmsLeft}{top\xspace}}{\providecommand{\cmsLeft}{left\xspace}}
\ifthenelse{\boolean{cms@external}}{\providecommand{\cmsRight}{bottom\xspace}}{\providecommand{\cmsRight}{right\xspace}}
\newcommand{\Acc}{\ensuremath{\mathcal{A}}\xspace}%
\newcommand{\AccEff}{\ensuremath{\mathcal{A}}\times\epsilon \xspace}%
\newcommand{\IECAL}    {I_{\text{ECAL}}}%
\newcommand{\IHCAL}    {I_{\text{HCAL}}}%
\newcommand{\ITRK}     {I_{\text{trk}}}%
\newcommand{\IRelComb} {I^{\text{rel}}_{\text{comb}}}%
\newcommand{\IRel}     {I^{\text{rel}}}%
\newcommand{\IRelmu}   {I^{\PGm}_{\text{rel}}}%
\newcommand{\IRele}    {I^{\Pe}_{\text{rel}}}%
\newcommand{\IComb}    {I_{\text{comb}}}%
\newcommand{\pp}{\ensuremath{{\Pp\Pp}}}%
\newcommand{\rts}{\ensuremath{\sqrt{s}}}%
\newcommand{\MN}{\ensuremath{\PGm\PGn}}%
\renewcommand{\EE}{\ensuremath{\Pep\Pem}}%
\renewcommand{\MM}{\ensuremath{\PGmp\PGmm}}%
\newcommand{\EN}{\Pe\PGn}%
\newcommand{\LN}{\ensuremath{\ell\PGn}}%
\newcommand{\MW}{\ensuremath{{m}_{\PW}}}%
\newcommand{\MZ}{\ensuremath{{m}_{\PZ}}}%
\providecommand{\MT}{\ensuremath{M_{\cmsSymbolFace{T}}}\xspace}%
\newcommand{\MLL}{\ensuremath{{M}_{\ell\ell}}}%

\newcommand{\Zll}{\ensuremath{\PZ \to \ell^+ \ell^-}}%
\newcommand{\Zee}{\ensuremath{\PZ \to \EE}}%
\newcommand{\Zmm}{\ensuremath{\PZ \to \MM}}%
\newcommand{\Ztt}{\ensuremath{\PZ \to \tau^+\tau^-}}%
\newcommand{\Wln}{\ensuremath{\PW \to \LN}}%
\newcommand{\Wen}{\ensuremath{\PW \to \EN}}%
\newcommand{\Wmn}{\ensuremath{\PW \to \MN}}%

\newcommand{\Wc}   {\ensuremath{\PW\hspace{-.12em}+\hspace{-.14em}\PQc}}
\newcommand{\PWmc} {\ensuremath{\PWm\hspace{-.12em}+\hspace{-.14em}\PQc}}
\renewcommand{\PWmp} {\ensuremath{\PWp\hspace{-.12em}+\hspace{-.14em}\PAQc}}
\newcommand{\gammac}   {\ensuremath{\PGg+\PQc}}
\newcommand{\Zc}   {\ensuremath{\PZ\hspace{-.12em}+\hspace{-.14em}\PQc}}
\newcommand{\Vc}   {\ensuremath{\mathrm{V}\hspace{-.12em}+\hspace{-.14em}\PQc}}
\newcommand{\Zcc}  {\ensuremath{\PZ+\PQc+\PAQc}}
\newcommand{\Zb}   {\ensuremath{\PZ\hspace{-.12em}+\hspace{-.14em}\PQb}}
\newcommand{\ZcZb} {\ensuremath{{(\PZ\hspace{-.12em}+\hspace{-.14em}\PQc)}\hspace{-.08em}/\hspace{-.14em}{(\PZ\hspace{-.12em}+\hspace{-.14em}\PQb)}}}
\newcommand{\Zuds} {\ensuremath{\PZ\hspace{-.12em}+\hspace{-.14em}\text{light flavour}}}
\newcommand{\Zj}   {\ensuremath{\PZ\hspace{-.12em}+\hspace{-.14em}\text{jets}}}
\newcommand{\Wj}   {\ensuremath{\PW\hspace{-.12em}+\hspace{-.14em}\text{jets}}}
\newcommand{\ZHF}   {\ensuremath{\PZ\hspace{-.12em}+\hspace{-.14em}\text{HF}}}
\newcommand{\HF}  {\text{HF}}
\newcommand{\parton}  {\text{parton}}
\newcommand{\genjet}  {\ensuremath{\,\text{gen jet}}\xspace}
\newcommand{\jet}  {\ensuremath{\,\text{jet}}\xspace}
\newcommand{\chadron}  {\PQc\text{ hadron}}
\newcommand{\bhadron}  {\PQb\text{ hadron}}
\newcommand{\chadrons}  {\PQc\text{ hadrons}}
\newcommand{\bhadrons}  {\PQb\text{ hadrons}}
\newcommand{\cjet}  {\PQc\text{ jet}}
\newcommand{\bjet}  {\PQb\text{ jet}}
\newcommand{\cjets}  {\PQc\text{ jets}}
\newcommand{\bjets}  {\PQb\text{ jets}}
\newcommand{\jets} {\ensuremath{\,\text{jets}}\xspace}
\newcommand{\SZc}  {\ensuremath{\sigma(\Zc)}}
\newcommand{\SZb}  {\ensuremath{\sigma(\Zb)}}
\newcommand{\ptZ} {\ensuremath{\pt^{\, \mathrm{Z}}}}
\newcommand{\SZcdiffpTZ} {\ensuremath{\frac{\rd\sigma(\Zc)}{\rd{\ptZ}} }}
\newcommand{\SZcdiffpTZl} {\ensuremath{\rd\sigma(\Zc)/\rd{\ptZ} }}
\newcommand{\SZbdiffpTZ} {\ensuremath{\frac{\rd\sigma(\Zb)}{\rd{\ptZ}} }}
\newcommand{\SZbdiffpTZl} {\ensuremath{\rd\sigma(\Zb)/\rd{\ptZ} }}
\newcommand{\SZcdiffpTjet} {\ensuremath{\frac{\rd\sigma(\Zc)}{\rd{\pt^{\jet}}} }}
\newcommand{\SZcdiffpTjetl} {\ensuremath{\rd\sigma(\Zc)/\rd{\pt^{\jet}} }}
\newcommand{\SZbdiffpTjet} {\ensuremath{\frac{\rd\sigma(\Zb)}{\rd{\pt^{\jet}}} }}
\newcommand{\SZbdiffpTjetl} {\ensuremath{\rd\sigma(\Zb)/\rd{\pt^{\jet}} }}
\newcommand{\ubar}{\PAQu}%
\newcommand{\dbar}{\PAQd}%
\newcommand{\sbar}{\PAQs}%
\newcommand{\cbar}{\PAQc}%
\newcommand{\tbar}{\PAQt}%
\newcommand{\ppZc}{\ensuremath{\Pp\Pp \to \PZ\hspace{-.12em}+\hspace{-.14em}\PQc \hspace{-.12em}+\hspace{-.14em}\PX}}%
\newcommand{\ppZb}{\ensuremath{\Pp\Pp \to \PZ\hspace{-.12em}+\hspace{-.14em}\PQb \hspace{-.12em}+\hspace{-.14em}\PX}}%
\newcommand{\ppZj}{\ensuremath{\Pp\Pp \to \PZ\hspace{-.12em}+\hspace{-.14em}\text{jets} \hspace{-.12em}+\hspace{-.14em}\PX}}%
\newcommand{\ppWc}{\ensuremath{\Pp\Pp \to \PW+\PQc + \PX}}%
\newcommand{\ppWpc}{\ensuremath{\Pp\Pp \to \PWp+\cbar}}%
\newcommand{\ppWmc}{\ensuremath{\Pp\Pp \to \PWm+\PQc}}%
\newcommand{\ppbarZc}{\ensuremath{\Pp\PAp \to \PZ+\PQc-\text{jet}}}%
\newcommand{\ppbarZb}{\ensuremath{\Pp\PAp \to \PZ+\PQb-\text{jet}}}%
\newcommand{\ppbarZj}{\ensuremath{\Pp\PAp \to \PZ+\text{jet}}}%

\newcommand{\ppZHtoqq}{\ensuremath{\Pp\Pp \to \PZ\hspace{-.12em}+\hspace{-.14em}\PH\hspace{-.12em}+\hspace{-.14em}\PX;\  \PH \to \PQq\PAQq}}%

\newcommand{\gcLO}{\ensuremath{\Pg\PQc \to \PZ \PQc}}%
\newcommand{\gcNLO}{\ensuremath{\Pg\PQc \to \PZ \PQc \Pg}}%
\newcommand{\qcNLO}{\ensuremath{\PQq\PQc \to \PZ \PQc \cPq}}%
\newcommand{\qqZcc}{\ensuremath{\PQq\PAQq \to \PZ\PQc\PAQc}}%

\newcommand{\Dpm}     {\ensuremath{{\PDpm}}}
\newcommand{\Dp}      {\ensuremath{{\PDp}}}
\newcommand{\Dm}      {\ensuremath{{\PDm}}}
\newcommand{\Dz}      {\ensuremath{{\PDz}}}
\newcommand{\Ds}      {\ensuremath{{\PDs}}}
\newcommand{\Lambdac} {\ensuremath{\PGLc}}
\newcommand{\Lambdab} {\ensuremath{\PGLb}}
\newcommand{\Dstar}   {\ensuremath{\PDstpm}}
\newcommand{\Dzbar}  {\ensuremath{\PADz}}
\newcommand{\Bpm}     {\ensuremath{\PBpm}}
\newcommand{\Dstarp} {\ensuremath{{\PDstp(2010)}}}
\newcommand{\Dstarm} {\ensuremath{{\PDstm(2010)}}}

\newcommand{\OSSS} {\ensuremath{\mathrm{OS}\hspace{-.12em}-\hspace{-.14em}\mathrm{SS}}}%

\newcommand{\ptmin}{\ensuremath{p_{\mathrm{Tmin}}}\xspace}
\newcommand{\ptmax}{\ensuremath{p_{\mathrm{Tmax}}}\xspace}

\newcommand{\rec}  {\text{rec}}

\newcommand{\PYTHIAsix} {{\PYTHIA{}6}\xspace}
\newcommand{\PYTHIAeight} {{\PYTHIA{}8}\xspace}
\newcommand{\aMCatNLO} {{MG5\_aMC}\xspace}

\newcommand\mean[1]{\left< #1 \right>}

\newcommand\T{\rule{0pt}{2.6ex}}
\newcommand\B{\rule[-1.2ex]{0pt}{0pt}}

\cmsNoteHeader{SMP-15-009}
\title{Measurement of associated Z + charm production in proton-proton collisions at $\sqrt{s} = 8$\TeV}
\titlerunning{Z + charm production in proton-proton collisions at $\sqrt{s} =  8$\TeV}

\date{\today}

\abstract{
A study of the associated production of a $\PZ$ boson and a charm quark jet ($\PZ + \PQc$), and a comparison to production with a $\PQb$ quark jet ($\PZ + \PQb$), in $\Pp\Pp$ collisions at a centre-of-mass energy of 8\TeV are presented. The analysis uses a data sample corresponding to an integrated luminosity of 19.7\fbinv, collected with the CMS detector at the CERN LHC. The $\PZ$ boson candidates are identified through their decays into pairs of electrons or muons. Jets originating from heavy flavour quarks are identified using semileptonic decays of $\PQc$ or $\PQb$ flavoured hadrons and hadronic decays of charm hadrons. The measurements are performed in the kinematic region with two leptons with $\pt^{\ell} > 20\GeV $, ${|\eta^{\ell}|} < 2.1$, $71 < m_{\ell\ell} < 111\GeV $, and heavy flavour jets with $\pt^{\text{jet}} > 25\GeV $ and ${|\eta^{ \text{jet}}|} < 2.5$. The $\PZ + \PQc$ production cross section is measured to be $\sigma(\Pp\Pp \to \PZ + \PQc + \PX) \mathcal{B}(\PZ \to \ell^+\ell^-) = 8.8 \pm 0.5\stat  \pm 0.6\syst \unit{pb}$. The ratio of the $\PZ + \PQc$ and $\PZ + \PQb$ production cross sections is measured to be $\sigma(\Pp\Pp \to \PZ + \PQc + \PX)/\sigma(\Pp\Pp \to \PZ + \PQb + \PX) = 2.0 \pm 0.2\stat \pm 0.2\syst$. The $\PZ + \PQc$ production cross section and the cross section ratio are also measured as a function of the transverse momentum of the $\PZ$ boson and of the heavy flavour jet. The measurements are compared with theoretical predictions.
}

\hypersetup{%
pdfauthor={CMS Collaboration},%
pdftitle={Measurement of associated Z + charm production in proton-proton collisions at sqrt(s) = 8 TeV},%
pdfsubject={CMS},%
pdfkeywords={CMS, physics, SMP, vector boson, heavy flavour, standard model physics}}

\maketitle

\section{Introduction}

The CERN Large Hadron Collider (LHC) has delivered a large sample of $\pp$ collisions containing events with a
vector boson (V) accompanied by one or more jets (V+jets).
Some of these events involve the production of a vector boson in association with jets originating from heavy flavour ($\HF$)
quarks and can be used to study specific predictions of the standard model (SM).

These V+jets events constitute an important background to many ongoing searches for new physics beyond the SM.
A proper characterization of these processes and validation of their theoretical description is important to provide a reliable estimate of their specific backgrounds to the various searches.
For example, third-generation scalar quarks (squarks) that are predicted by supersymmetric theories to decay via charm quarks have been searched for in final states with a charm quark jet ($\cjet$) and a large transverse momentum imbalance~\cite{Aad:2014nra, Aad:2015gna, Khachatryan:2015wza}.
A dominant background to this process is the associated production of a $\cjet$ and a $\PZ$ boson that decays invisibly into neutrinos.
An improved description of this background can be obtained from a measurement of the same process with the $\PZ$ boson decaying into charged leptons.

Similarly, the associated production of a $\PZ$ boson and $\HF$ jets is a significant background to the production of the Higgs boson in association with a $\PZ$ boson
($\ppZHtoqq$). Experimental studies of this process in the context of the SM focus on an analysis with $\PQb$ quarks in the final
state~\cite{Chatrchyan:2012ww, Chatrchyan:2013zna, Aad:2012gxa, Aad:2014xzb},
although some models beyond the SM also predict enhanced decay rates in the $\PQc\PAQc$ final state~\cite{Delaunay:1621790}.
In either case, it is important to understand the relative contribution of the different flavours to the
$\ZHF$ jets background to minimize the associated systematic uncertainties.

The possibility of observing evidence of an intrinsic charm (IC) quark component in the nucleon has recently
received renewed interest~\cite{Brodsky:2015fna}. The associated production of neutral vector bosons and $\cjets$ ($\Vc$) has been
identified~\cite{Beauchemin:2014rya, Bailas:2015jlc, Lipatov:2016feu, Boettcher:2015sqn} as a suitable process to investigate this physics topic.
One of the main effects of an IC component would be an enhancement of $\Zc$ production, mainly at large values of the transverse momentum of the $\PZ$ boson
and of the $\cjet$.

Production of a $\PZ$ boson and a $\cjet$ has been studied in high-energy hadron collisions by the D0~\cite{D0-Zc} and CDF~\cite{CDF-VD} experiments
at the Tevatron $\Pp\PAp$ collider. More recently, the LHCb Collaboration has measured the associated production of a $\PZ$ boson and a $\PD$ meson in the forward
region in $\Pp\Pp$ collisions at $\sqrt{s} =7 \TeV$~\cite{LHCB-ZD}.

In this paper we present a measurement of the production cross section at $\sqrt{s} =8\TeV$ of a $\PZ$ boson and at least one jet from a $\PQc$ quark.
In addition, the relative production of a $\PZ$ boson and a jet from heavy quarks of different flavours ($\PQc$ or $\PQb$) is quantified by the ratio of
their production cross sections.
The associated production of a $\PZ$ boson and at least one or two $\bjets$ using an inclusive $\PQb$ tagging technique to identify $\Zb$ events has been studied with the same
dataset and the results are reported in Ref.~\cite{Khachatryan:2016iob}. To reduce the uncertainties in the ratio, the production cross section of a $\PZ$ boson
and a jet from a $\PQb$ quark is remeasured in this analysis using exactly the same methodology as for the $\Zc$ cross section.
The remeasured $\Zb$ cross section agrees with the published value within one standard deviation and is used in the ratio measurement.

The $\PZ$ boson is identified through its decay into a pair of electrons or muons.
Jets with $\HF$ quark content are identified through (1) the semileptonic decay of $\PQc$ or $\PQb$ flavoured hadrons with a muon in the
final state, and (2) using exclusive hadronic decays of charm hadrons.
The cross section and cross section ratio are measured at the level of stable particles, which are defined prior to the emission of any electroweak radiation.
To minimize acceptance corrections, the measurements are restricted to a phase space
that is close to the experimental fiducial volume with optimized sensitivity for the investigated processes:
two leptons with transverse momentum $\pt^{\ell}>20\GeV$, pseudorapidity $|\eta^{\ell}| < 2.1$, and
dilepton invariant mass consistent with the mass of the $\PZ$ boson, $71 < m_{\ell\ell} < 111\GeV$,
together with a $\PQc$ ($\PQb$) jet with $\pt^{\jet} > 25\GeV$, $|\eta^{\jet}| < 2.5$. The jet should be separated from the leptons of the $\PZ$ boson candidate
by a distance $\Delta R ({\text{jet}},\ell) = \sqrt{\smash[b]{(\Delta\eta)^2 +(\Delta\phi)^2}} > 0.5$.
The cross section $\sigma(\ppZc)\mathcal{B}(\PZ \to \ell^+\ell^-)$
(abbreviated as $\SZc\,\mathcal{B}$) and the cross section ratio $\sigma(\ppZc)/\sigma(\ppZb)$ (abbreviated as $\SZc/\SZb$) are determined
both inclusively and differentially as a function of the transverse momentum of the $\PZ$ boson, $\ptZ$, and the $\pt$ of the jet
with heavy flavour content, $\pt^{\jet}$.

The paper is structured as follows. The CMS detector is briefly described in Section~\ref{sec:CMS_det}, and the data and simulated samples used are
presented in Section~\ref{sec:samples}. Section~\ref{sec:event_sel} deals with the selection of the $\ZHF$ jets signal sample, the
auxiliary samples of events from the associated production of $\Wc$, and top quark-antiquark ($\ttbar$) production.
The determination of the $\PQc$ tagging efficiency is the subject of Section~\ref{sec:ctag}.
The analysis strategy devised to separate the two contributions, $\Zc$ and $\Zb$, in the sample of $\ZHF$ jets is detailed in Section~\ref{sec:sigext}.
Section~\ref{sec:syst_uncert} reviews the most important sources of systematic uncertainties and their impact on the measurements.
Finally, the measurements of the inclusive $\Zc$ cross section and the $\ZcZb$ cross section ratio
are presented in Section~\ref{sec:xsec_inc}, and the differential measurements are reported in Section~\ref{sec:xsec_diff}.
The main results of the paper are summarized in Section~\ref{sec:summary}.

\section{The CMS detector~\label{sec:CMS_det}}
The central feature of the CMS apparatus is a superconducting solenoid
of 6\unit{m} internal diameter, providing a magnetic field of 3.8\unit{T}.
Within the solenoid volume are a silicon pixel
and strip tracker, a lead tungstate crystal electromagnetic calorimeter (ECAL),
and a brass and scintillator hadron calorimeter,
each composed of a barrel and two endcap sections.
Extensive forward calorimetry complements the coverage
provided by the barrel and endcap detectors.
The silicon tracker measures charged particles within the pseudorapidity
range $|\eta|< 2.5$. It consists of 1440 silicon pixel
and 15\,148 silicon strip detector modules.
For nonisolated particles of $1 < \pt < 10\GeV$ and $|\eta| < 1.4$, the
track resolutions are typically 1.5\% in $\pt$ and 25--90 (45--150)\mum
in the transverse (longitudinal) impact parameter~\cite{CMS-PAPER-TRK-11-001}.
The electron momentum is estimated by combining the energy measurement
in the ECAL with the momentum measurement in the tracker. The momentum
resolution for electrons with $\pt \approx 45\GeV$ from $\Zee$ decays
ranges from 1.7\% for nonshowering electrons in the barrel region to 4.5\% for
showering electrons in the endcaps~\cite{CMS-PAPER-EGM-13-001}.
Muons are measured in the pseudorapidity range $ |\eta|< 2.4$,
using three technologies: drift tubes,
cathode strip chambers, and resistive plate chambers.
Matching muons to tracks measured in the silicon tracker
results in a relative transverse momentum resolution for muons
with $20 < \pt < 100\GeV$ of 1.3--2.0\% in the barrel and
better than 6\% in the endcaps.
The \pt resolution in the barrel is better than 10\% for muons with \pt up to 1\TeV~\cite{CMS-PAPER-MUO-10-004}.
For nonisolated muons with $1 < \pt < 25 \GeV$, the relative transverse momentum resolution is 1.2--1.7\% in the barrel and 2.5--4.0\%
in the endcaps~\cite{CMS-PAPER-TRK-11-001}.
The first level of the CMS trigger system~\cite{Khachatryan:2212926}, composed of custom hardware processors, uses information from the calorimeters and muon detectors to select events of interest in a fixed time interval of less than 4\mus. The high-level trigger processor farm further decreases the event rate from around 100\unit{kHz} to less than 1\unit{kHz}, before data storage.
A more detailed description of the CMS detector, together with a definition of the coordinate system used and the basic kinematic variables, can be found in Ref.~\cite{Chatrchyan:2008zzk}.

\section{Data and simulated samples~\label{sec:samples}}

The data were collected by the CMS experiment during 2012 at the $\pp$ centre-of-mass energy of 8\TeV
and correspond to an integrated luminosity of $\mathcal{L}= 19.7 \pm 0.5\fbinv$.

Samples of simulated events are produced with Monte Carlo (MC) event generators, both for the signal process and for the main
backgrounds. A sample of signal $\PZ$ boson events is generated with \MADGRAPH v5.1.3.30~\cite{Alwall:2011uj},
interfaced with \PYTHIA v6.4.26~\cite{Pythia6} for parton showering and hadronization using the MLM~\cite{Alwall:2007fs, Alwall:2008qv} matching scheme.
The \MADGRAPH generator produces parton-level events with a vector boson and up to four partons at leading order (LO) on the basis of a matrix-element calculation.
The generation uses the parton distribution functions (PDF) set CTEQ6L~\cite{Pumplin:2002vw}.
The matching scale between jets from matrix element calculations and those produced via parton showers is 10 GeV,
and the factorization and renormalization scales are set to $q^2 =  M^2_{\PZ}+{(\ptZ)}{}^2$.

Other physics processes produce events with the same final state topology as the signal.
The main background is the production of $\ttbar$ events.
Smaller contributions are expected from the direct production of a pair of vector bosons: $\PW\PW$, $\PW\PZ$, and $\PZ\PZ$.

{\tolerance=1200
A sample of $\ttbar$ events is generated with \POWHEG v1.0~\cite{Campbell:2014kua, Nason:2004rx, Frixione:2007vw, Alioli:2010xd}, interfaced with \PYTHIAsix and using the CT10~\cite{Gao:2013xoa} PDF set.
The $\PW\PW$, $\PW\PZ$, and $\PZ\PZ$ processes are modelled with samples of events generated with \PYTHIAsix and the CTEQ6L1 PDF set.
\par}

A sample of $\PW$ boson events is generated with \MADGRAPH interfaced with \PYTHIAsix.
It is used in the determination of the $\PQc$ tagging efficiency and to validate
the modelling of relevant distributions with a data sample of $\Wj$ events.
The matching scale between jets from matrix element calculations and those produced via parton showers is 10 GeV, and the factorization
and renormalization scales are set to $q^2 =  M^2_{\PW}+{(\pt^{\mathrm{W}})}{}^2$.
For all event generation the \PYTHIAsix parameters for the underlying event modelling are set to the Z2$^{\ast}$ tune~\cite{Chatrchyan:2013gfi}.

{\tolerance=700
Generated events are processed through a full \GEANTfour-based~\cite{Agostinelli:2002hh} CMS detector simulation and trigger emulation.
Simulated events are then reconstructed using the same algorithms as used to reconstruct collision data
and are normalized to the integrated luminosity of the data sample using their respective cross sections.
For electroweak processes the cross sections are evaluated to next-to-next-to-leading order (NNLO) with \FEWZ3.1~\cite{Li:2012wna},
using the MSTW2008NNLO~\cite{Martin:2009iq} PDF set.
The cross sections for diboson production are evaluated at next-to-leading order (NLO) with \MCFM 6.6~\cite{Campbell:2010ff} and using the MSTW2008NLO~\cite{Martin:2009iq} PDF set.
The $\ttbar$ cross section is taken at NNLO from Ref.~\cite{Czakon:2013goa}.
The simulated samples incorporate additional $\pp$ interactions in the same or neighbouring bunch crossings (pileup).
Simulated events are weighted so that the pileup distribution
matches the measured one, with an average of about 21 pileup interactions per bunch crossing.
\par}

Simulated samples are corrected for differences between data and MC descriptions of lepton trigger, reconstruction, and selection efficiencies ($\epsilon_{\ell}$).
Lepton efficiencies are evaluated with samples of dilepton events in the $\PZ$ mass peak with the ``tag-and-probe'' method~\cite{CMS-PAPER-EWK-10-005}, and
correction factors $\epsilon_{\ell}^\text{data}/\epsilon_{\ell}^{\mathrm\mathrm{MC}}$, binned in terms of $\pt$ and $\eta$
of the leptons, are computed.
These correction factors, based on the kinematics of each lepton in an event, are multiplied and used as an event weight.

The simulated signal sample includes $\PZ$ boson events accompanied by jets originating from quarks of all flavours ($\PQb$, $\PQc$, and light).
Events are classified as $\Zb$, $\Zc$, or $\Zuds$ according to the flavour of the generator-level jets
built from all showered particles after fragmentation and hadronization (all stable particles
except neutrinos) and clustered with the same algorithm that is used to reconstruct data jets.
A generator-level jet is defined to be $\PQb$ flavoured if $\pt^{\genjet} > 15\GeV$ and there is a $\PQb$ hadron among
the particles generated in the event within a cone of radius $\Delta R = 0.5$
around the jet axis. Similarly, a generator-level jet is considered to be $\PQc$ flavoured if $\pt^{\genjet} > 15\GeV$ and there is a $\PQc$ hadron and
no $\PQb$ hadrons within a cone of $\Delta R = 0.5$ around the jet axis.
A $\Zj$ event is assigned as a $\Zb$ event if there is at least a generator-level jet identified as a $\PQb$ flavoured jet
regardless of the number of $\PQc$ flavoured or light jets, $\Zc$ if there is at least a $\PQc$ flavoured jet at the
generator-level and no $\PQb$ flavoured generator-level jets, and $\Zuds$ otherwise.

\section{Event reconstruction and selection~\label{sec:event_sel}}

Electron and muon candidates are reconstructed following standard CMS procedures~\cite{CMS-PAPER-MUO-10-004, CMS-PAPER-EGM-13-001}.
Jets, missing transverse energy, and related quantities are determined using the CMS particle-flow (PF) reconstruction algorithm~\cite{Sirunyan:2017ulk},
which identifies and reconstructs stable particle candidates arising from a collision with an optimized combination of the signals measured from all subdetectors.

Jets are built from PF candidates using the anti-$\kt$ clustering algorithm~\cite{Cacciari:2008gp} with a distance parameter of $R = 0.5$.
The energy and momentum of the jets are corrected as a function of the jet $\pt$ and $\eta$ to account for the nonlinear response of the
calorimeters and for the presence of pileup interactions~\cite{CMS-PAPER-JME-10-011, CMS-PAPER-JME-13-004}.
Jet energy corrections are derived using samples of simulated events and further adjusted using dijet, photon+jet and $\PZ$+jet events in data.

The missing transverse momentum vector $\ptvecmiss$ is the projection on the plane perpendicular to the beams of the negative vector sum of the momenta
of all particles that are reconstructed with the PF algorithm. The missing transverse energy variable, $\ETmiss$, is defined as the magnitude of the $\ptvecmiss$
vector, and it is a measure of the transverse energy of particles leaving the detector undetected~\cite{CMS-PAS-JME-12-002}.

The primary vertex of the event, representing the hard interaction, is selected among the reconstructed vertices as the one with the highest sum of the transverse momenta squared of the tracks associated to it.

\subsection{Selection of \texorpdfstring{$\ZHF$}{Z + HF} jet events~\label{sec:Zsel}}

Events with a pair of leptons are selected online by a trigger system that requires the presence of two lepton candidates of the same flavour with
$\pt > 17$ and $8\GeV$ for the leading-$\pt$ and subleading-$\pt$ lepton candidates, respectively.
The analysis follows the offline selections as used in the CMS $\Zee$ and $\Zmm$ inclusive analyses~\cite{CMS-PAPER-EWK-10-005} and requires the presence of
two high-$\pt$ reconstructed leptons with opposite charges in the pseudorapidity region $\abs{\eta^{\ell}} < 2.1$. The transverse momentum of the leptons has to be greater than $20\GeV$.

The leptons are required to be isolated. The combined isolation $\IComb$ is used to quantify the additional hadronic activity around the selected leptons.
It is defined as the sum of the transverse energy of neutral hadrons and photons and the transverse momentum of charged particles in a cone with
$R < 0.3$ $(0.4)$ around the electron (muon) candidate, excluding the contribution from the lepton itself.
Only charged particles originating from the primary vertex are considered in the sum to minimize the contribution from pileup interactions.
The contribution of neutral particles from pileup vertices is estimated and subtracted from $\IComb$.
For electrons, this contribution is evaluated with the jet area method described in Ref.~\cite{jet_area}; for muons,
it is taken to be half the sum of the $\pt$ of all charged particles in the cone originating from pileup vertices.
The factor one-half accounts for the expected ratio of charged to neutral particle energy in hadronic interactions.
The electron (muon) candidate is considered to be isolated when $\IComb/\pt^{\ell} < 0.15$ $(0.20)$.
Finally, the analysis is restricted to events with a dilepton invariant mass, $m_{\ell\ell}$, in the range $91\pm 20\GeV$ in accordance with previous $\Zj$ measurements~\cite{Khachatryan:2016iob, Khachatryan:2016crw}.

A $\Zj$ sample is selected by requiring the presence of at least one jet with $\pt^{\jet}>25\GeV$ and $|\eta^{\jet}|<2.5$.
Jets with an angular separation between the jet axis and any of the selected leptons less than $\Delta R ({\text{jet}},\ell) = 0.5$ are not considered.
To reduce the contribution from $\ttbar$ events, we require $\ETmiss$ to be smaller than $40\GeV$.

{\tolerance=700
Hadrons with $\PQc$ or $\PQb$ quark content decay weakly with lifetimes of the order of $10^{-12}\unit{s}$ and mean decay lengths larger than 100\unit{$\mu$m} at the LHC energies.
Secondary vertices well separated from the primary vertex can be reconstructed from the tracks of their charged decay products.
We focus on the following three signatures to identify jets originating from a heavy flavour quark:
\begin{itemize}
\item[$\centerdot$] \textbf{ {Semileptonic mode}} --- A semileptonic decay of a heavy flavour hadron leading to a well-identified muon associated to a displaced secondary vertex.
\item[$\centerdot$] \textbf{ {$\PDpm$ mode}} --- A displaced secondary vertex with three tracks consistent with a $\PDpm \to \PK^\mp \PGp^\pm\PGp^\pm$ decay.
\item[$\centerdot$] \textbf{ {$\Dstar$ mode}} --- A displaced secondary vertex with two tracks consistent with a
$\Dz\to \PKm\PGp^+$ ($\Dzbar\to \PKp\PGp^-$) decay and associated with a $\Dstarp\to\Dz \PGp^+$ ($\Dstarm\to\Dzbar \PGp^-$)
decay at the primary vertex.
\end{itemize}
Displaced secondary vertices for the first two categories are formed with either the Simple Secondary Vertex (SSV)~\cite{CMS-PAPER-BTV-12-001} or the
Inclusive Vertex Finder (IVF)~\cite{Khachatryan:2011wq, Chatrchyan:2013zja} CMS vertex reconstruction algorithms.
Both algorithms follow the adaptive vertex fitter technique~\cite{Adaptive_Vertex} to  construct a secondary vertex, but differ in the tracks used.
The SSV algorithm takes as input the tracks constituting the jet; the IVF algorithm starts from a displaced track with respect to the primary vertex
(\textit{seed} track) and searches for nearby tracks, in terms of their separation distance in three dimensions and their angular separation
around this \textit{seed}, to build the vertex. Tracks used in a secondary vertex reconstruction must have $\pt>1\GeV$.
Vertices reconstructed with the IVF algorithm are considered first because of the higher efficiency of the algorithm. If no IVF vertex is found, SSV vertices are
searched for, thus providing additional event candidates.
We employ a different technique for the third ($\Dstar$ mode) category, as described below in the text. The typical mass resolution in the $\PDpm$ and $\Dstar$ reconstruction is ${\approx}17\MeV$ in the decay modes analyzed here.
\par}

\subsubsection{Selection in the semileptonic mode}

The $\Zc$ ($\Zb$) events with a semileptonic $\PQc$ ($\PQb$) quark decay are selected by looking for a reconstructed muon
(\textit{muon-inside-a-jet}) among the constituents of any of the selected jets.
This \textit{muon-inside-a-jet} candidate has to satisfy the same quality criteria as those imposed on the muons from the $\PZ$ boson decay.
The muon has to be reconstructed in the region $\abs{\eta^{\PGm}} < 2.4$, with $\pt^{\PGm}<25\GeV$, $\pt^{\PGm}/\pt^{\jet}<0.6$,
and it should not be isolated from hadron activity. The combined isolation has to be large, $\IComb/\pt^{\PGm} >0.2$.
Furthermore, the \textit{muon-inside-a-jet} is required to be associated to a secondary vertex, reconstructed either with the IVF or SSV algorithm.
No minimum $\pt$ is required for the muon beyond the general $\pt>1\GeV$ requirement for the tracks used in the reconstruction of the secondary vertices.
Muon reconstruction sets a natural threshold of $\pt \gtrapprox 3\GeV$ in the barrel region and $\pt \gtrapprox 2\GeV$ in the endcaps to ensure
the muon passes the material in front of the muon detector and travels deep enough into the muon system to be reconstructed and satisfy the identification
criteria~\cite{CMS-PAPER-EWK-10-005}.
The above selection results in 4145 events in the $\Zee$ channel and 5258 events in the $\Zmm$ channel.

\begin{figure*}[!tb]
\begin{center}
\includegraphics[width=\cmsDoubleFigWidth]{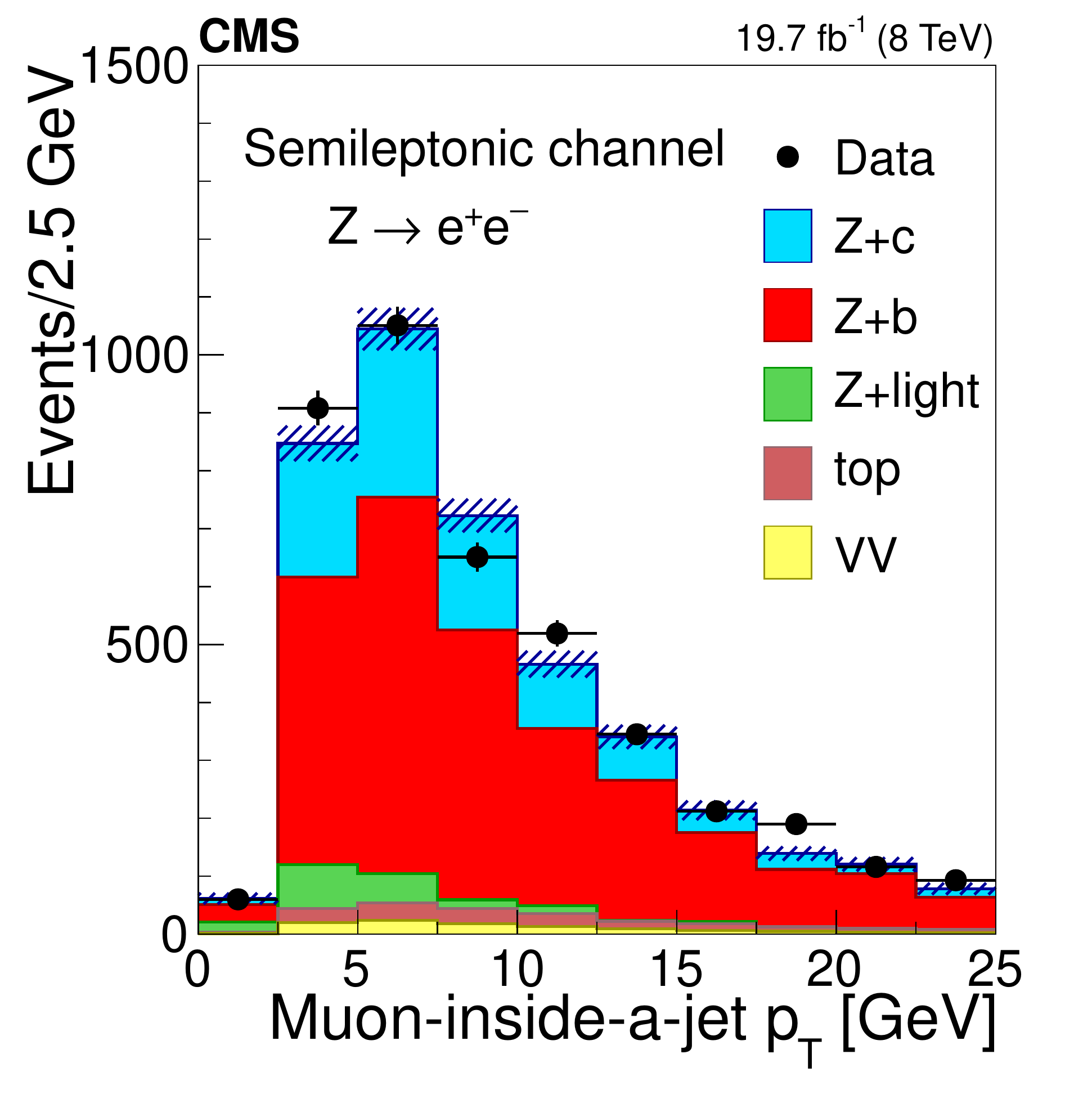}
\includegraphics[width=\cmsDoubleFigWidth]{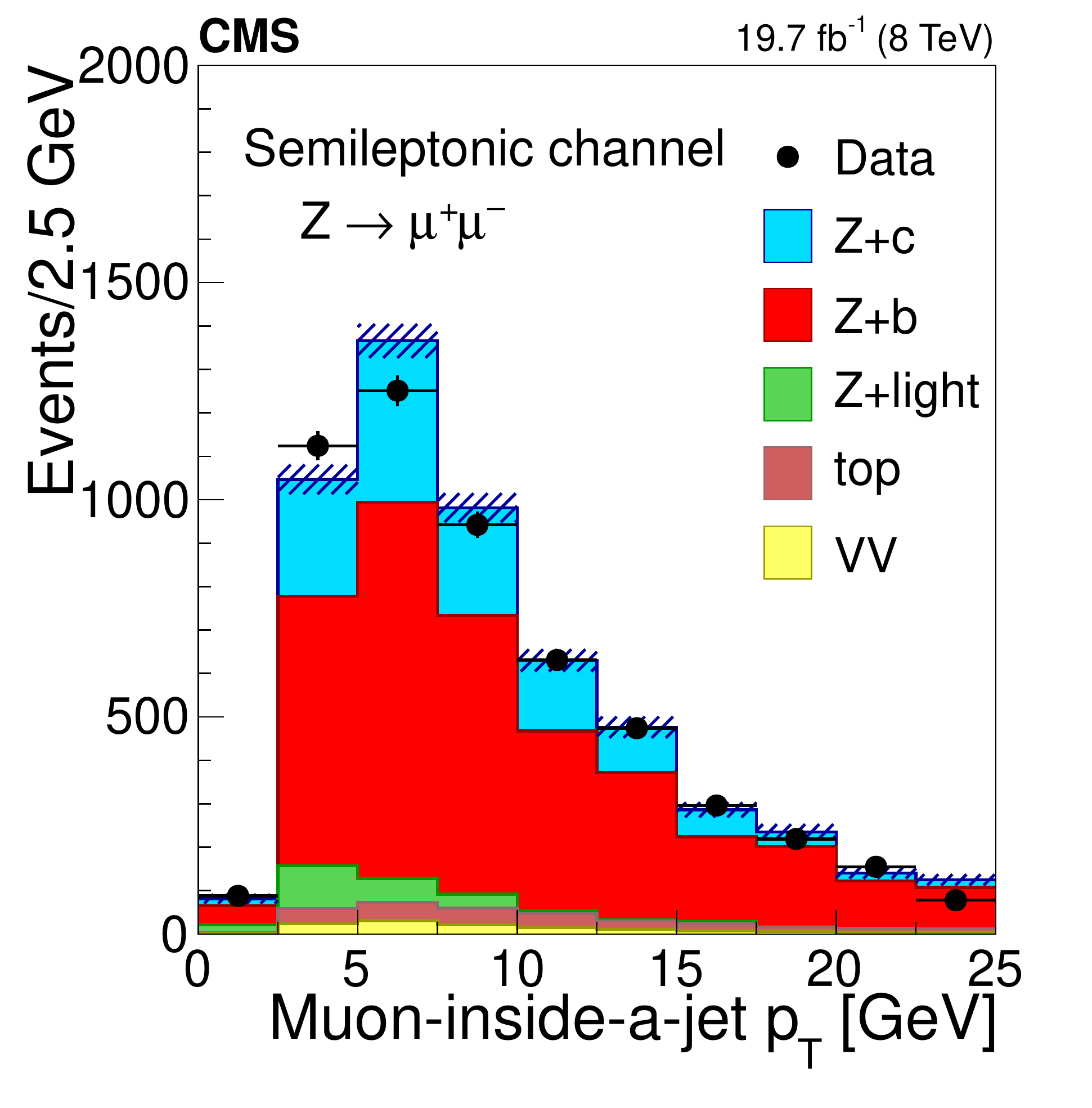}
\caption{Transverse momentum distribution of the selected \textit{muon-inside-a-jet} for events with an identified muon among the jet constituents,
in the dielectron (left) and dimuon (right) channels.
The contributions from all processes are estimated with the simulated samples.
Vertical bars on data points represent the statistical uncertainty in the data. The hatched areas represent the statistical uncertainty in the MC simulation.
}
\label{fig:muon_in_jet_pt}
\end{center}
\end{figure*}

Figure~\ref{fig:muon_in_jet_pt} shows the transverse momentum distribution of the selected \textit{muon-inside-a-jet}
for $\Zee$ (left) and $\Zmm$ (right). The data are compared with the predictions of the MC simulations,
which are composed of $\Zb$ events (${\approx}65\%$), $\Zc$ events (${\approx}25\%$),
$\Zuds$ (${\lesssim}5\%$), and other backgrounds, such as $\ttbar$ and diboson production (${\approx}5\%$).

\subsubsection{Selection in the $\Dpm$ mode}

Event candidates in the $\PDpm$ mode are selected by looking for secondary vertices made of three tracks and with a reconstructed
invariant mass consistent with the $\Dpm$ mass: $1869.5 \pm 0.4 \MeV$~\cite{PDG}.
The sum of the charges of the tracks participating in the secondary vertex must be ${\pm}1$.
The kaon mass is assigned to the track with opposite sign to the total charge of the three-prong vertex, and the remaining tracks are assumed to have the mass of a charged pion. This assignment is correct in more than 99\% of the cases, since the fraction of double Cabibbo-suppressed decays is extremely small~\cite{PDG}.

\begin{figure*}[t]
\begin{center}
\includegraphics[width=\cmsDoubleFigWidth]{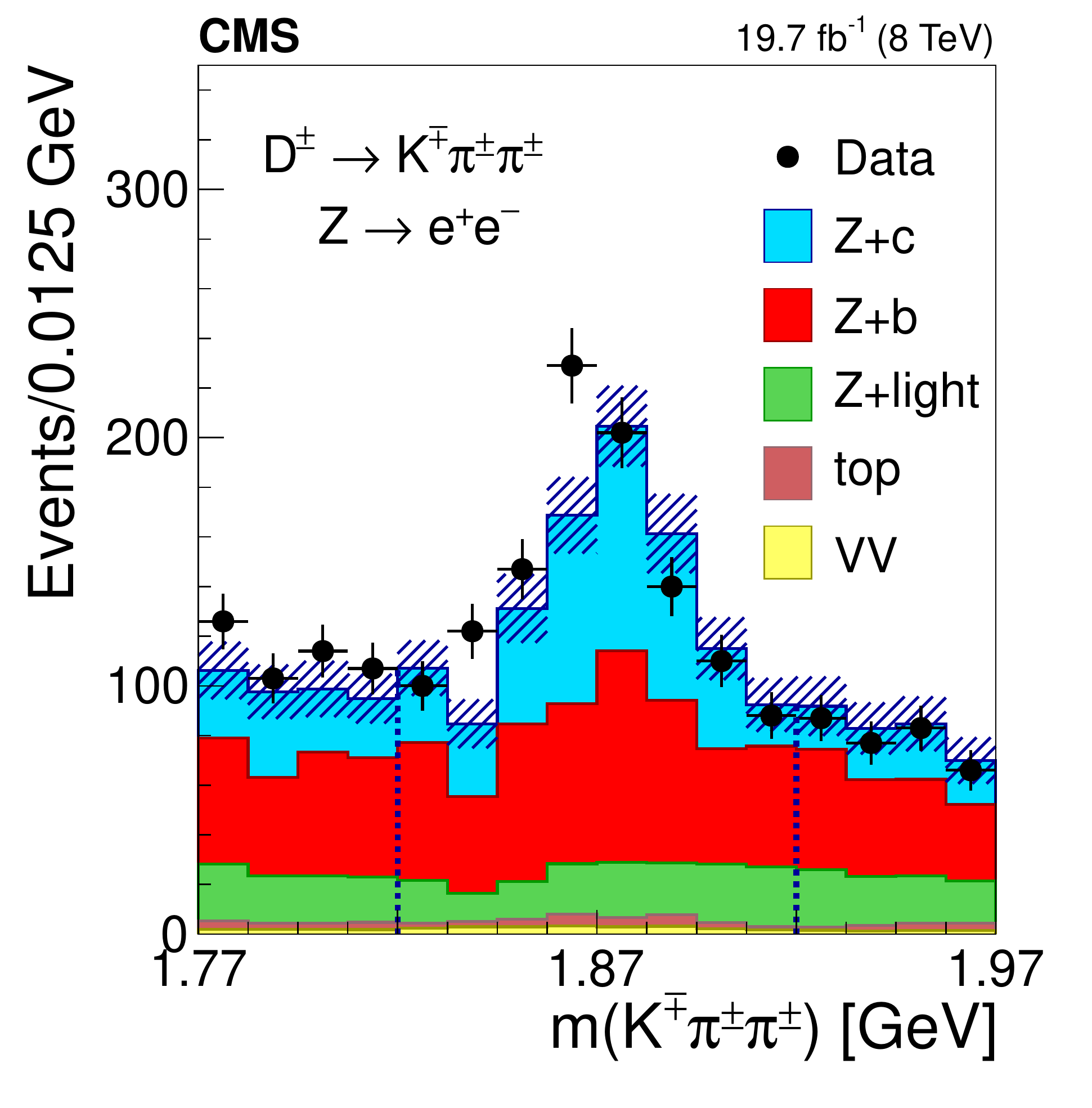}
\includegraphics[width=\cmsDoubleFigWidth]{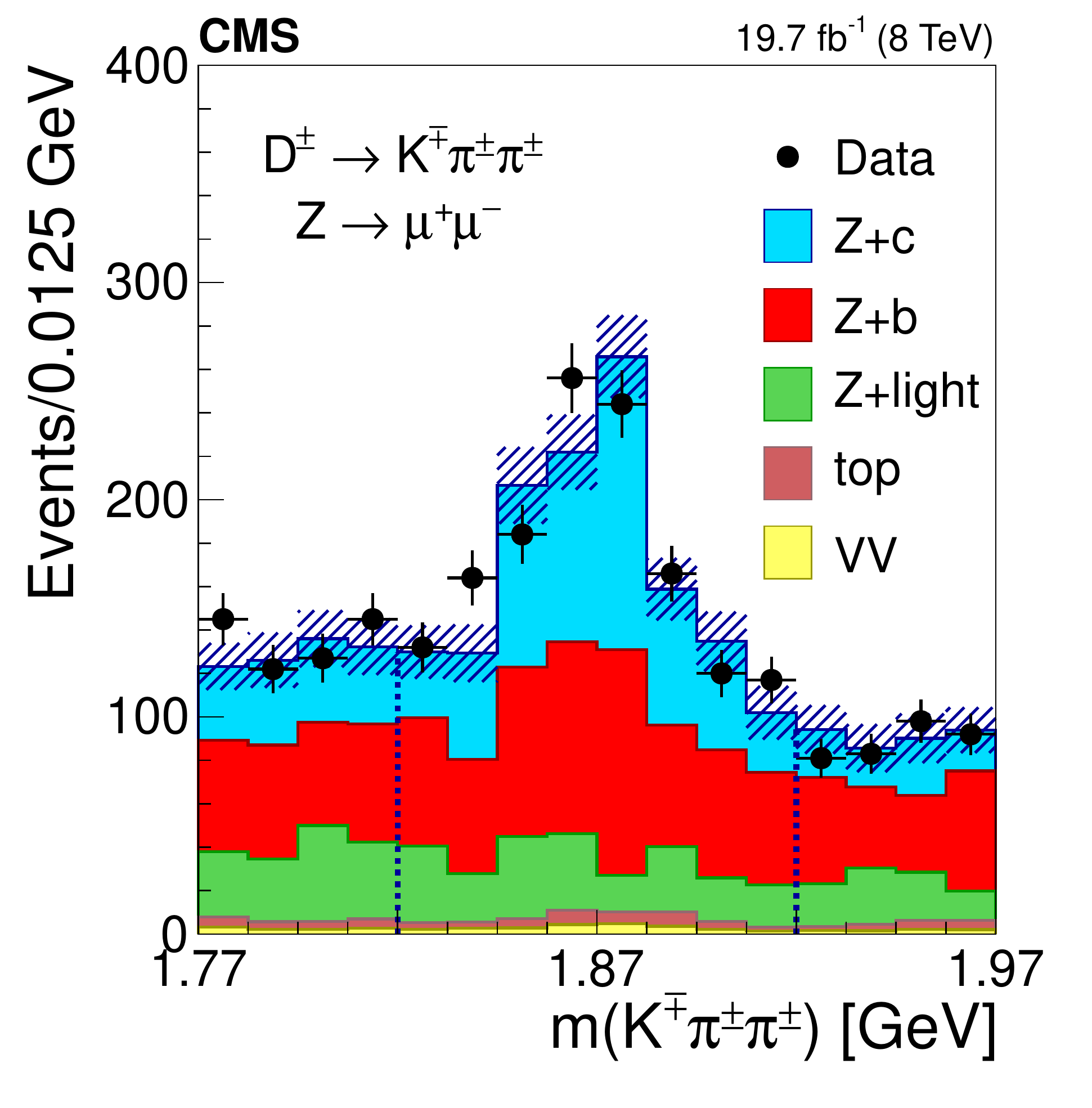}
\caption{The invariant mass distribution of three-prong secondary vertices for events selected in the $\Dpm$ mode, in the dielectron (left) and dimuon (right) channels.
The mass assigned to each of the three tracks is explained in the text. The contributions from all processes are estimated with the simulated samples.
The two dashed, vertical lines indicate the mass range of the signal region.
Vertical bars on data points represent the statistical uncertainty in the data. The hatched areas represent the statistical uncertainty in the MC simulation.
}
\label{fig:mass_Dpm}
\end{center}
\end{figure*}

The distribution of the reconstructed invariant mass for $\Dpm$ candidates associated with $\Zee$ (left) and $\Zmm$ (right) is presented in Fig.~\ref{fig:mass_Dpm}.
The signal and background contributions shown in the figure are estimated with the simulated samples.
The charm fraction $\mathcal{B}(\PQc\to\Dpm)$ in the \PYTHIA simulation ($19.44 \pm 0.02)\%$ is lower than the value ($22.7 \pm 0.9 \pm 0.5)\%$ obtained from a
combination~\cite{gladilin} of published measurements performed at LEP~\cite{OPAL,ALEPH,DELPHI} and the branching fraction of the
decay $\Dpm \to \PK^\mp \PGp^\pm\PGp^\pm$ ($7.96 \pm 0.03)\%$, is also lower than the PDG value ($9.13 \pm 0.19)\%$~\cite{PDG};
predicted event rates from the MC simulation are scaled in order to match the experimental charm fractions.

{\tolerance=10000
The signal region is defined by the constraint $\Delta m(\Dpm) \equiv |m^{\rec}(\Dpm)-1.87\GeV|<0.05\GeV$, where $m^{\rec}(\Dpm)$ is the reconstructed
mass of the $\Dpm$ meson candidate.
The mass range of the signal region is indicated in Fig.~\ref{fig:mass_Dpm} as two dashed, vertical lines.
The width of the signal region approximately corresponds to three times the measured mass resolution.
The nonresonant background is subtracted from the event count in the signal window, and is estimated using the number of events selected in a control region away from the resonance,
extending up to a window of $0.1\GeV$ width, $N[0.05<\Delta m(\Dpm)<0.10\GeV]$, as also shown in Fig.~\ref{fig:mass_Dpm}.
\par}

The number of selected events in data after background subtraction is $375 \pm 44$ in the $\Zee$ channel and $490 \pm 48$ in the $\Zmm$ channel.
Based on the simulation, the selected sample is enriched in $\Zc$ events (${\approx}60\%$), while the fraction of $\Zb$ events is ${\approx}35\%$.
The contribution from $\Zuds$ events is negligible, and the contribution of $\ttbar$ and diboson events is smaller than 5\%.

\subsubsection{Selection in the $\Dstar$ mode}

\begin{figure*}[t]
\begin{center}
\includegraphics[width=\cmsDoubleFigWidth]{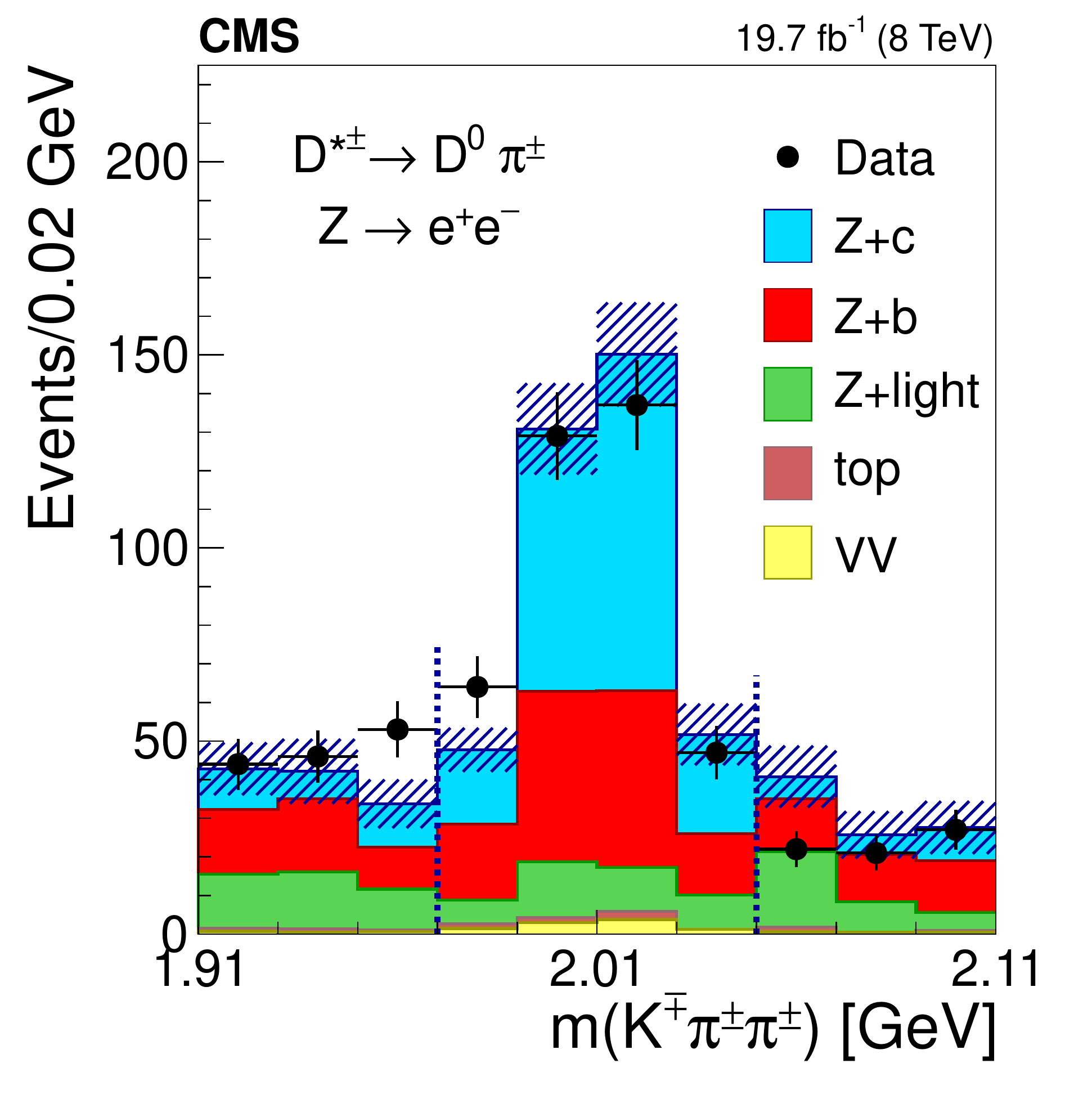}
\includegraphics[width=\cmsDoubleFigWidth]{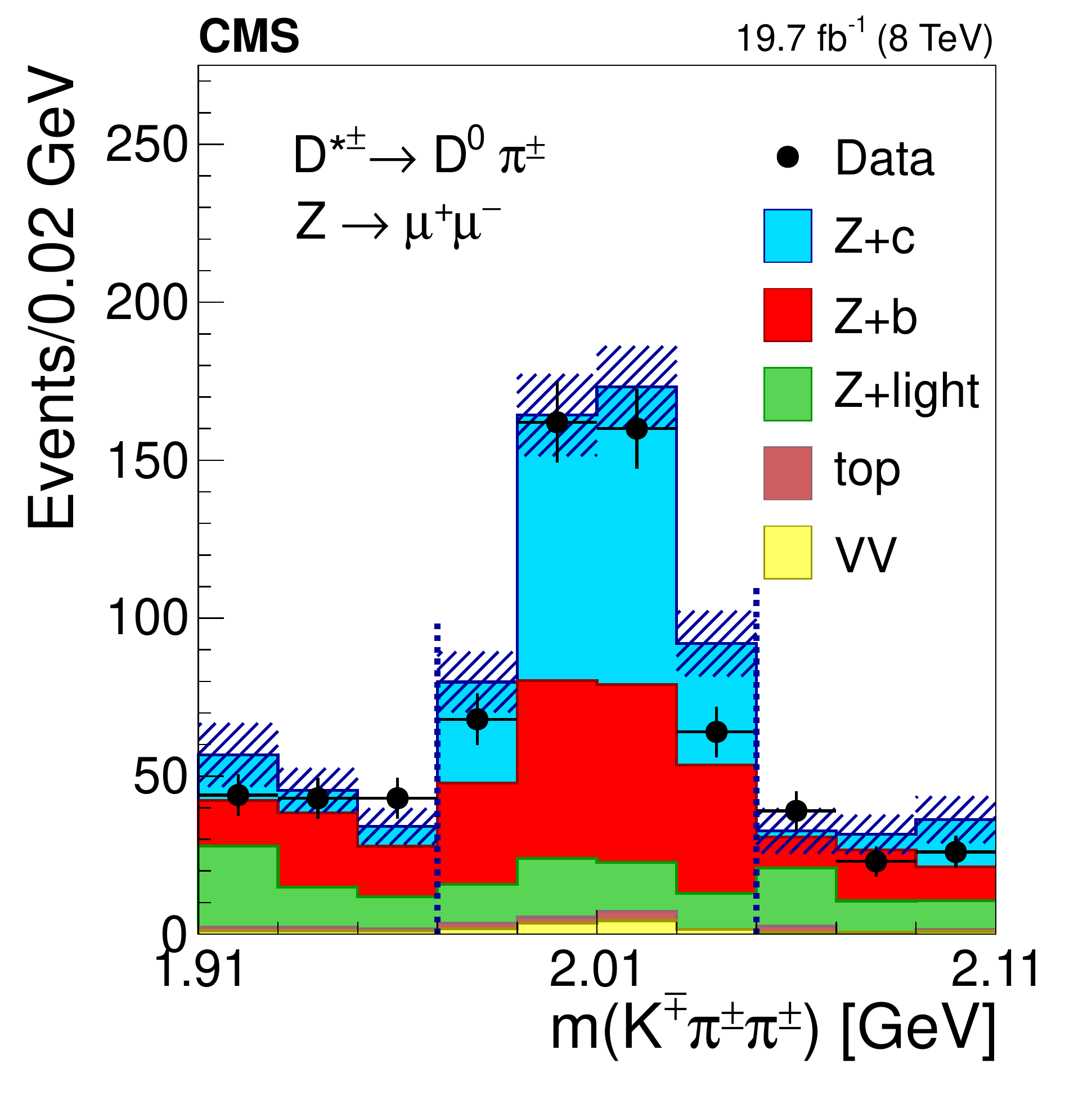}
\caption{The invariant mass distribution of the three-track system composed of a two-prong secondary vertex and a primary particle for events selected in the $\Dstar$ mode,
in the dielectron (left) and dimuon (right) channels.
The mass assigned to each of the three tracks is explained in the text. The contributions from all processes are estimated with the simulated samples.
The two dashed, vertical lines mark the mass range of the signal region.
Vertical bars on data points represent the statistical uncertainty in the data. The hatched areas represent the statistical uncertainty in the MC simulation.
}
\label{fig:mass_Dstar}
\end{center}
\end{figure*}

Events with $\Zj$ candidates in the  $\Dstar$  mode are selected by requiring a displaced vertex with two oppositely charged tracks among the tracks constituting the jet.
These tracks are assumed to be the decay products of a $\Dz$ meson.
The candidate is combined with a third track from the jet constituents that should represent the \textit{soft pion}, emitted in the
strong decay $\Dstarp\to\Dz \PGp^+$.
To be a \textit{soft pion} candidate, the track must have a transverse momentum larger than $0.5\GeV$ and lie in a cone of radius $\Delta R (\Dz,\PGp)=0.1$ around the line of flight of the $\Dz$ meson candidate.

The track of the $\Dz$ meson candidate with a charge opposite to the charge of the \textit{soft pion} is taken to be the
kaon from the $\Dz$ meson decay and is required to have $\pt>1.75\GeV$.
The other track is assigned to be the pion and is required to have $\pt>0.75\GeV$.
Two-track combinations with an invariant mass different from the nominal $\Dz$ meson mass (1864.86 $\pm$ 0.13\MeV) by less than 100\MeV are selected, and a
secondary vertex is constructed using the two tracks and the CMS Kalman vertex fitter algorithm~\cite{VertexFitter}.
The two-track system is kept as a valid $\Dz$ meson candidate if the probability for the vertex fit is greater than 0.05.

To ensure a clean separation between the secondary and primary vertices, the 2D-distance in the transverse plane between them,
divided by the uncertainty in the distance measurement (defined as decay length significance) has to be larger than 3.
Furthermore, to guarantee that the reconstructed vertex corresponds to a two-body decay of a
hadron originating at the primary vertex, the momentum vector of the $\Dz$ meson candidate has to be collinear
with the line from the primary vertex to the secondary vertex: the cosine of the angle between the two directions has to be larger than 0.99.
Finally, only events with a mass difference between the $\Dstar$ and $\Dz$ candidates within 5\MeV
from the expected value ($145.426 \pm  0.002\MeV$~\cite{PDG}) are selected.

{\tolerance=3000
The product of the branching fractions
$\mathcal{B}(\PQc\to\Dstarp) \mathcal{B}(\Dstarp\to\Dz\PGp^+) \mathcal{B}(\Dz\to \PKm \PGp^+)$ (+ charge conjugate)
in the \PYTHIA simulation is $(0.741 \pm 0.005)\%$, which is about 15\% larger than
the average of the experimental values, $(0.622 \pm  0.020)\%$~\cite{gladilin, PDG}.
Therefore, expected event rates from the MC simulation are scaled in order to match the experimental values.
\par}

{\tolerance=1200
The distribution of the reconstructed mass of the $\Dstar$ candidates is presented in Fig.~\ref{fig:mass_Dstar} for events with a $\PZ$ boson decaying into $\Pe^+\Pe^-$
(left) and $\PGm^+\PGm^-$ (right). The contribution from the different processes is estimated with the simulated samples.
\par}

{\tolerance=10000
The signal region is defined by the constraint $\Delta m(\Dstar) \equiv |m^{\rec}(\Dstar)-2.01\GeV| < 0.04\GeV$, where $m^{\rec}(\Dstar)$
is the reconstructed mass of the $\Dstar$ candidate, and which corresponds to slightly more than twice the measured mass resolution.
The two dashed, vertical lines present in Fig.~\ref{fig:mass_Dstar} indicate the mass range of the signal region.
The nonresonant background contribution to the signal region is subtracted using the number of events selected in a control region away from the resonance.
We use a window of $0.12~(2\times 0.06)\GeV$ width, $N[0.04<\Delta m(\Dstar)<0.10\GeV]$, also shown in Fig.~\ref{fig:mass_Dstar}, and apply the proper weight
to account for the different width of the signal and control regions $(8/12)$.
\par}

The number of data selected events after background subtraction is $234 \pm 22$ in the $\Zee$ channel and $308 \pm 24$ in the $\Zmm$ channel.
According to the predictions obtained from the simulated samples, the fraction of $\Zc$ events in the selected sample is high (${\approx}65\%$) and the contribution of $\Zb$ events is ${\approx}30\%$. No contribution is expected from $\Zuds$ events. Less than 5\% of the selected events arise from $\ttbar$ and diboson production.

Systematic biases due to the background subtraction are expected to be negligible compared to the statistical uncertainty, because of the approximate agreement
observed between data and simulation as shown in Figs.~\ref{fig:mass_Dpm} and~\ref{fig:mass_Dstar}.

\subsection{Selection of \texorpdfstring{$\PW$}{W}+charm jet events (\texorpdfstring{\PQc}{c} jet control sample)~\label{sec:Wsel}}

Additional data and simulated samples consist of events from associated production of a $\PW$ boson and a jet originating from a $\PQc$ quark ($\Wc$).
They are used to model characteristic distributions of jets with $\PQc$ quark content and to measure the $\PQc$ tagging efficiency in a large, independent sample.
Jet flavour assignment in the simulated $\Wj$ events follows the criteria presented in Section~\ref{sec:samples} for $\Zj$ events.

The production of a $\PW$ boson in association with a $\PQc$ quark proceeds at LO via the processes $\PQq\Pg \to \PWmc$ and
$\PAQq \Pg \to \PWmp$ ($\PQq = \PQs, \PQd$).
A key property of the $\PQq\Pg\to \PW+\PQc$ reaction is the presence of a charm quark and a W boson with opposite-sign (OS) charges.
Background processes deliver OS and same-sign (SS) events in equal proportions, whereas $\PQq\Pg\to \PW+\PQc$ is always OS.
Therefore, distributions obtained after $\OSSS$ subtraction are representative of the $\Wc$ component, allowing for detailed studies of $\cjets$.

{\tolerance=700
We select $\Wc$ events following the criteria of the analysis reported in Ref.~\cite{WplusC}.
Candidate events are selected online using single-lepton triggers, which require at least one isolated electron (muon) with $\pt > 27~(24)\GeV$ and $\abs{\eta^{\ell}} < 2.1$.
The lepton identification and isolation criteria are very similar to those used for the $\Zj$ selection.
The offline $\pt$ threshold is increased to 30~(25)\GeV for electrons (muons) because of the higher thresholds of the single-lepton triggers.
The transverse invariant mass of the lepton and $\ptvecmiss$ system is defined as $\MT = \sqrt{\smash[b]{2\: \pt^{\ell}\: \ETmiss \: [1-\cos(\phi^{\ell}-\phi^{\ETmiss})]}}$,
where $\phi^{\ell}$ and $\phi^{\ETmiss}$ are the azimuthal angles of the lepton momentum and $\ptvecmiss$.
The $\MT$ must be larger than 55 (50)\GeV for events in the $\Wen$ ($\Wmn$) channel.
\par}

Identification of jets originating from $\PQc$ quarks proceeds exactly as described in Section~\ref{sec:Zsel}.
In all cases the charge of the $\PQc$ quark is unequivocally known. In the semileptonic mode the charge of the muon determines the charge of the $\PQc$ quark.
In the $\Dpm$ and $\Dstar$ modes the charge of the $\PD$ candidates defines the charge of the $\PQc$ quark.
OS events are events when the muon, $\Dpm$, or $\Dstar$ candidate has a charge opposite to the lepton
from the $\PW$ boson decay, and SS events when the charge is the same.

Based on the simulations, after subtracting the SS from the OS samples, $\Wc$ events are the dominant contributor to the distributions;
${\approx}90\%$ in the semileptonic decay modes and larger than 98\% in the $\Dpm$ and $\Dstar$ exclusive channels.
The remaining backgrounds, mainly from top quark production, are subtracted using the simulation.

\subsection{Selection of \texorpdfstring{\ttbar}{ttbar} samples~\label{sec:ttbar_sel}}

A sample of $\ttbar$ events ($\Pe\PGm$-$\ttbar$ sample) is selected using the leptonic decay modes of the $\PW$ bosons from the $\ttbar$ pair when they
decay into leptons of different flavour.
The $\ttbar$ production is a natural source of $\PQb$ flavoured jets and enables tests of the MC description of the relevant distributions for $\bjets$ as well as
of the performance of the secondary vertexing method.
This sample is also used to model the $\ttbar$ background in the discriminant variables used to extract the signal yields.

An $\Pe\PGm$-$\ttbar$ sample is selected online by a trigger path based on the presence of an electron-muon pair.
The offline selection proceeds as for the $\ZHF$ jet events, but the two leptons must be different flavours.
After the selection, contributions from processes other than $\ttbar$ production are negligible.

An additional $\ttbar$ enriched sample is used to estimate the normalization of the remaining $\ttbar$ background.
The same selection used for the $\ZHF$ jet signal is applied: two leptons of the same flavour, $\Pe\Pe$ or $\PGm\PGm$,
and $\ETmiss > 80\GeV$, instead of $\ETmiss < 40\GeV$.
The small contribution from $\Zj$ events in these samples $({\lesssim}3\%)$ is subtracted according to its MC expectation.

\section{Measurement of the \texorpdfstring{$\PQc$ and $\PQb$}{c and b} quark tagging efficiencies~\label{sec:ctag}}

The accuracy of the description in the MC simulations of the secondary vertex reconstruction part of the $\PQc$ tagging method is evaluated with a control sample of $\Wc$ events with a
well-identified \textit{muon-inside-a-jet}.
The events are selected as described in Section~\ref{sec:Wsel} except for the requirement that the \textit{muon-inside-a-jet} must come from a secondary vertex.
The $\OSSS$ strategy suppresses all backgrounds to the $\Wc$ sample in the $\Wmn$ decay mode except for Drell--Yan events.
The contamination from the Drell--Yan process, which yields genuine OS dimuon events may reach 25\%.
The $\Wc$ sample in the $\Wen$ decay mode, with the lepton from the $\PW$ decay of different flavour from the \textit{muon-inside-a-jet}, is not affected by
this background and is employed for the $\PQc$ tagging study.

A $\Wc$ event is ``SV-tagged'' if there is a reconstructed secondary vertex in the jet and the \textit{muon-inside-a-jet} is one of the tracks
used to form the vertex. The $\cjet$ tagging efficiency is the fraction of ``SV-tagged'' $\Wc$ events, over all $\Wc$ events, after $\OSSS$ subtraction:
\begin{equation*}
\epsilon_{\PQc}=\frac{N(\Wc)^{\OSSS}(\text {SV-tagged})}{N(\Wc)^{\OSSS}}.
\end{equation*}

Efficiencies are obtained independently with the data and with the $\Wj$ simulated samples.
Data-to-simulation scale factors, $SF_{\PQc}$, are then computed as the ratio between the $\cjet$ tagging efficiencies in data and simulation,
\begin{equation*}
SF_{\PQc}=\frac{\epsilon^{\text{data}}_{\PQc}}{\epsilon^{\mathrm{MC}}_{\PQc}}.
\end{equation*}
They are used to correct the simulation efficiency.

The $\cjet$ tagging efficiencies and the scale factors are computed both inclusively and as a function of the jet $\pt$.
The expected average $\PQc$ tagging efficiency is ${\approx}33\%$ for the IVF algorithm and ${\approx}21\%$ for the SSV algorithm.
The $\PQc$ tagging efficiency ranges from 24\% for the IVF algorithm (15\% for the SSV algorithm) for $\pt^{\jets}$ of 25--30 \GeV and up to 37\% (26\%) for $\pt^{\jets}$ of $\approx$100\GeV.
The $SF_{\PQc}$ for jets with a $\pt$ larger than $25\GeV$ is found to be $0.93\pm 0.03 \stat \pm 0.02 \syst$ for IVF vertices.
It is $0.92\pm 0.03 \stat \pm 0.02 \syst$ for SSV vertices.
The systematic uncertainty accounts for inaccuracies in pileup description, jet energy scale and resolution, lepton efficiencies, background subtraction, and modelling of charm production and decay fractions in the simulation.

\begin{figure*}[!htb]
\begin{center}
\includegraphics[width=\cmsDoubleFigWidth]{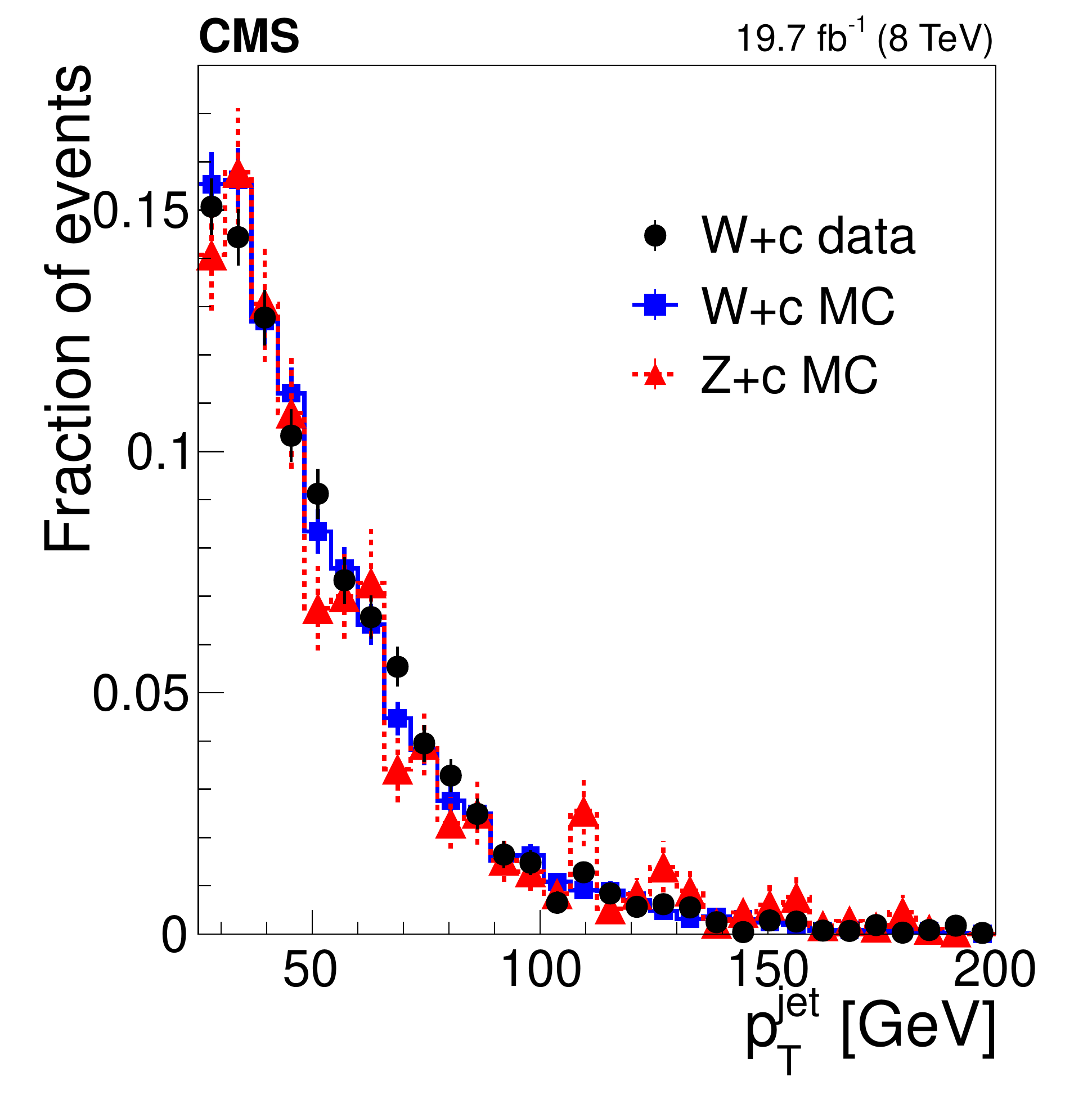}
\includegraphics[width=\cmsDoubleFigWidth]{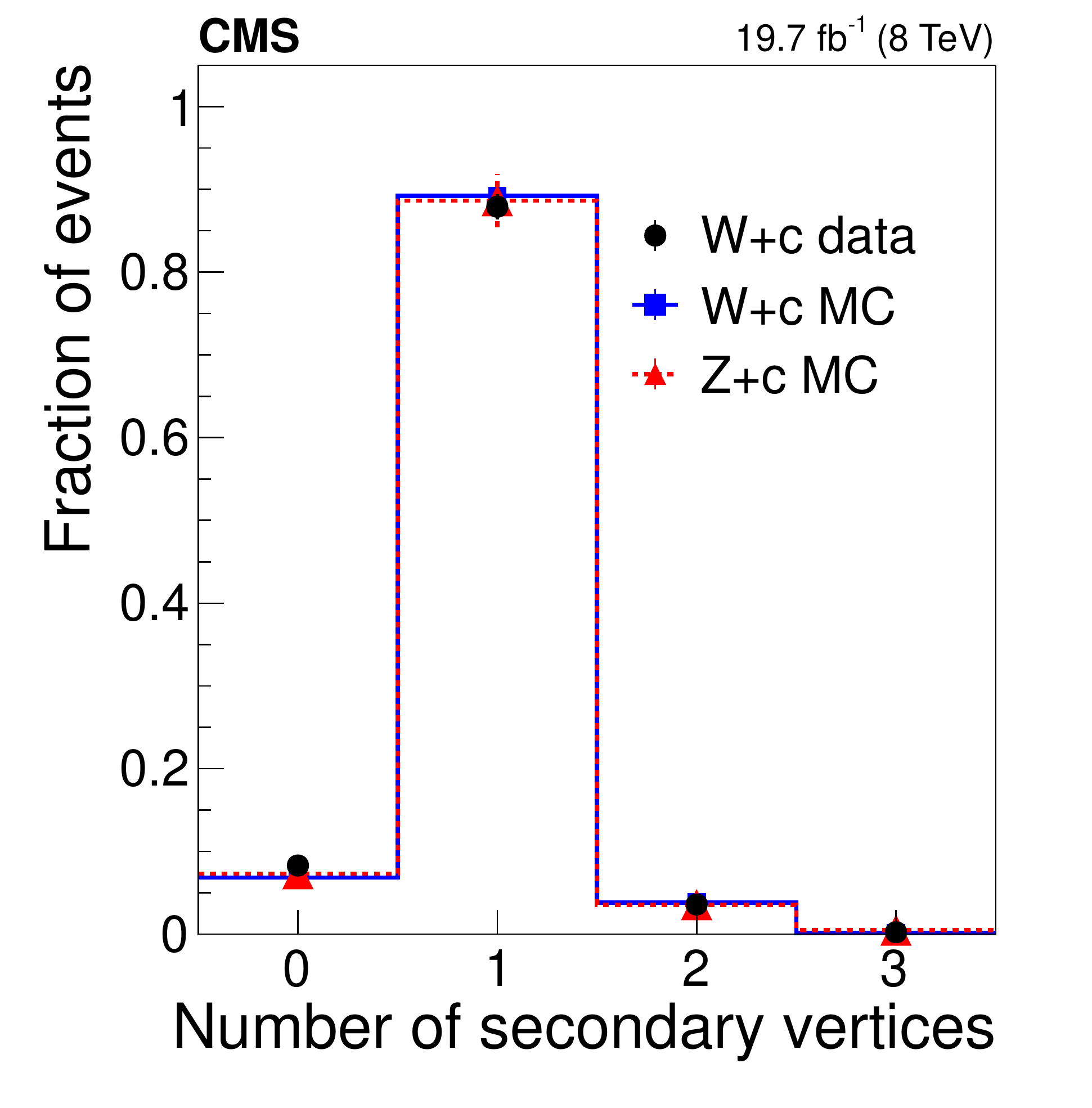}
\caption{Transverse momentum distribution of the $\PQc$-tagged jet (left) and number of reconstructed secondary vertices (right), normalized to unity, in simulated
$\Wc$ and $\Zc$ samples and in $\Wc$ data events. The $\Wc$ distributions are presented after the $\OSSS$ subtraction.
Vertical bars represent the statistical uncertainties.
}
\label{fig:Wc_Zc}
\end{center}
\end{figure*}

Detailed studies of the behaviour of the $\PQb$ tagging methods developed in CMS are available in Ref.~\cite{CMS-PAS-BTV-13-001}.
Following the same procedure, we have used the $\Pe\PGm$-$\ttbar$ sample to investigate the data-to-MC agreement for the $\PQb$ tagging methods in this analysis.
The $\PQb$ tagging efficiencies in data and simulated events are computed as the fraction of $\Pe\PGm$-$\ttbar$ events with a \textit{muon-inside-a-jet} participating in a secondary vertex with respect to the number of events when the secondary vertex condition is released.
The $SF_{\PQb}=\epsilon^{\text{data}}_{\PQb}/\epsilon^{{\rm MC}}_{\PQb}$ is measured to be $0.96 \pm 0.03$ for both IVF and SSV vertices,
where the uncertainty includes statistical and systematic effects due to the jet energy scale and resolution and the pileup.

\section{Analysis strategy~\label{sec:sigext} }

The extraction of $\Zc$ and $\Zb$ event yields is based on template fits to distributions of variables sensitive to the jet flavour.
In the semileptonic mode we use the corrected invariant mass, $M_\text{vertex}^\text{corr}$ (corrected secondary-vertex mass),
of the charged particles attached to the secondary vertex (the \textit{muon-inside-a-jet} included).
All charged particles are assigned the mass of the pion, except for the identified muon.
A correction is included to account for additional particles, either charged or neutral,
that may have been produced in the semileptonic decay but were not reconstructed~\cite{Aaij:2015yqa},
\begin{equation*}
 M_\text{vertex}^\text{corr} = \sqrt{M^2_\text{vertex} + p^2_\text{vertex} \sin^2 \theta}  + p_\text{vertex} \sin \theta,
\end{equation*}
where $M_\text{vertex}$ and $p_\text{vertex}$ are the invariant mass and modulus of the vectorial sum of the momenta of all reconstructed particles
associated to the secondary vertex, and $\theta$ is the angle between the momentum vector sum and the vector from the primary to the secondary
vertex.

In the $\Dpm$ and $\Dstar$ modes a likelihood estimate of the probability that the jet tracks come from the primary vertex,
called jet probability (JP) discriminant~\cite{CMS-PAPER-BTV-12-001}, is used.

The shapes of the $\Zc$ discriminant distributions are modelled in data using OS $\Wc$ events, after subtraction of the
SS $\Wc$ distributions. It is checked using simulated events that the corresponding distributions obtained from the $\Wc$ samples accurately describe the
$\Zc$ distributions.
The main features of the jets, such as $\pt$, $\eta$, jet charged multiplicity, and the
number of secondary vertices are found to be consistent between $\Zc$ and $\Wc$ simulated samples and are in agreement with the observed distributions
in the sample of $\Wc$ events in data.
Figure~\ref{fig:Wc_Zc} (left) shows the simulated $\pt^{\jet}$ distributions of $\Wc$ and $\Zc$ events compared to $\Wc$ data after $\OSSS$ subtraction.
The number of secondary vertices, identified with the IVF algorithm, is shown in Fig.~\ref{fig:Wc_Zc} (right). Events with no reconstructed IVF vertices have at least one reconstructed vertex with the SSV vertex algorithm. All distributions in Fig.~\ref{fig:Wc_Zc} are normalized to unity.
\begin{figure*}[!htb]
\begin{center}
\includegraphics[width=0.3\textwidth]{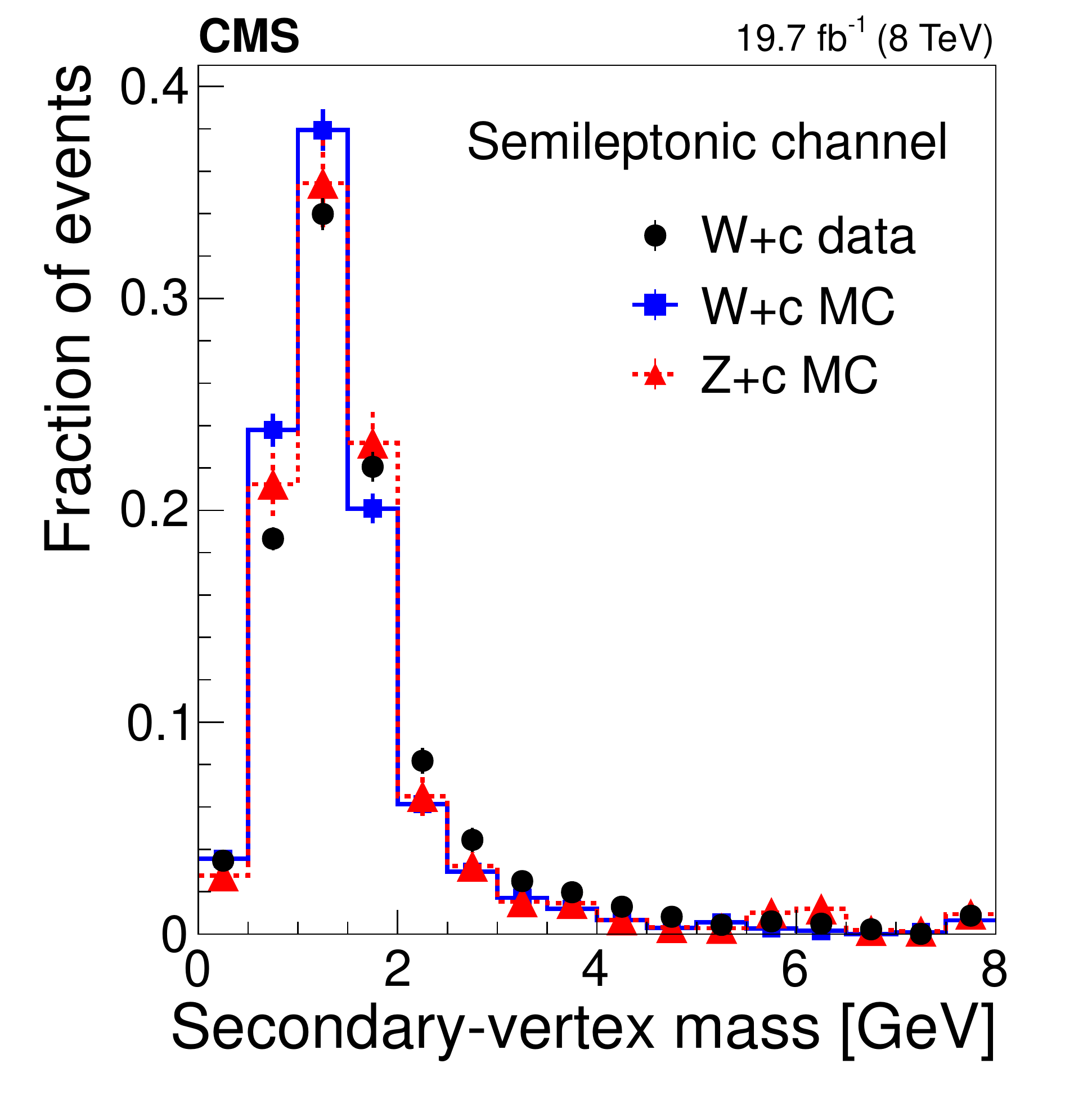}
\includegraphics[width=0.3\textwidth]{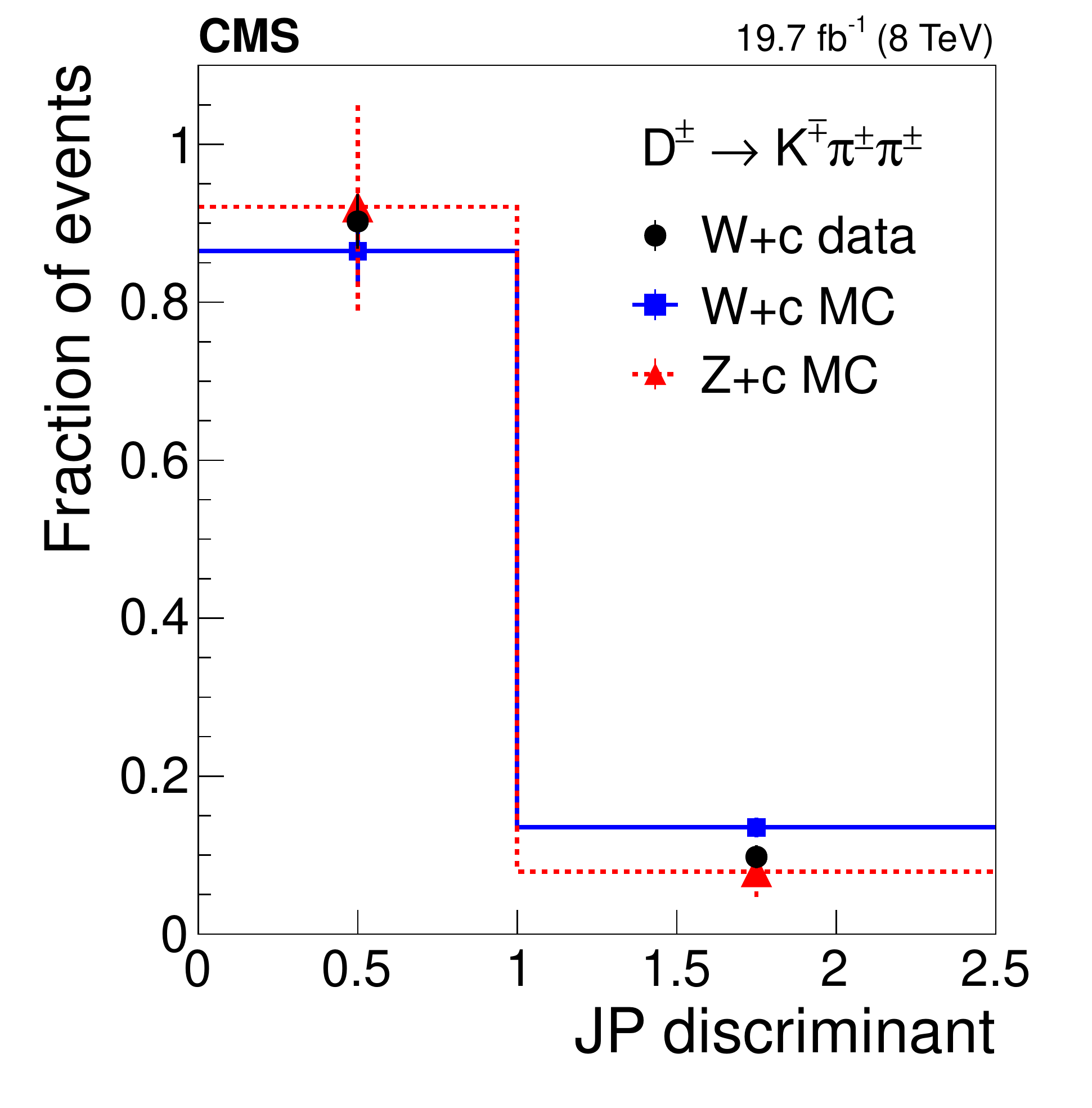}
\includegraphics[width=0.3\textwidth]{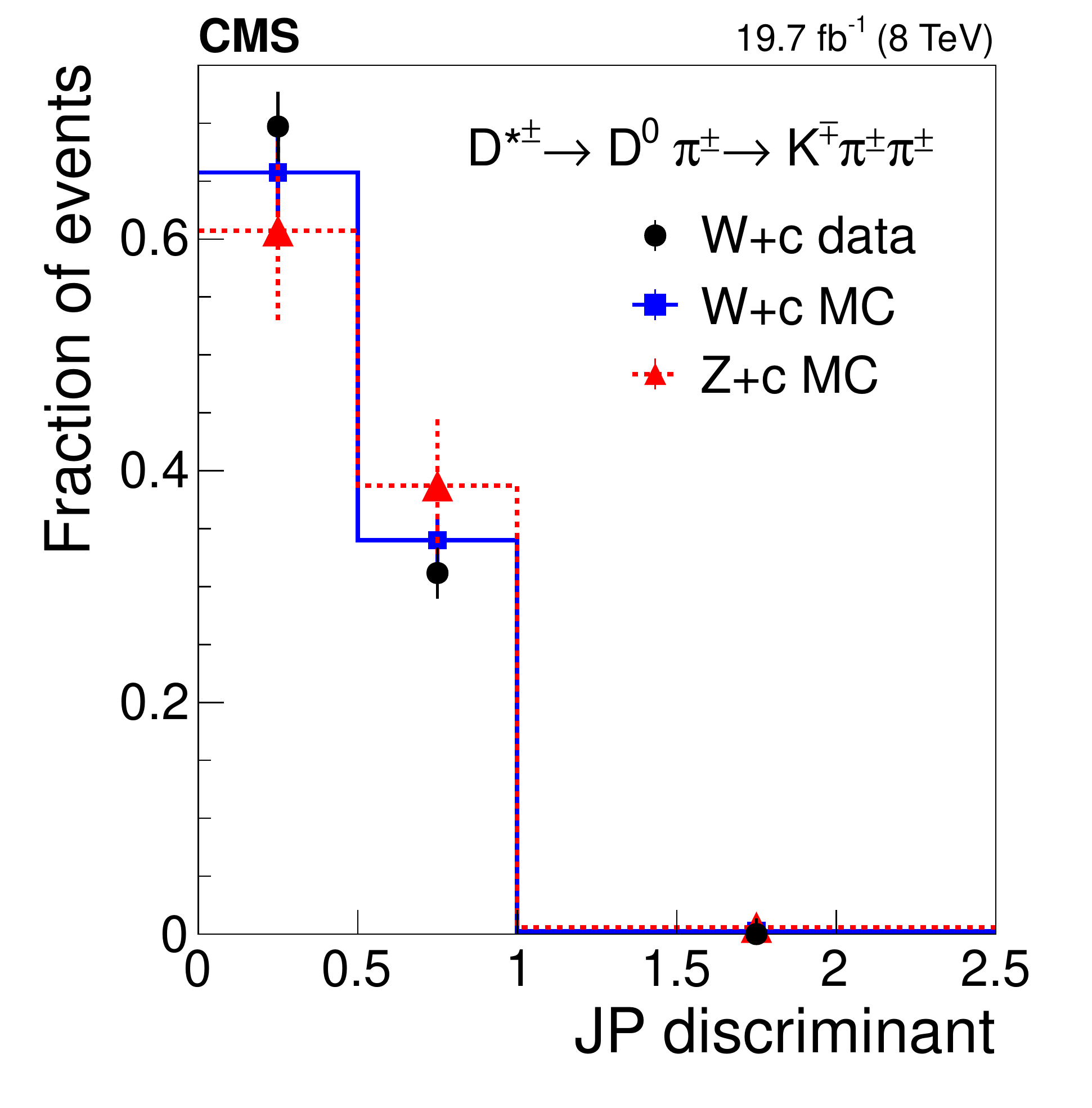}
\caption{Distributions of the corrected secondary-vertex mass (left plot) and JP discriminant ($\Dpm$ and $\Dstar$ modes in the middle and right plots), normalized to unity, in simulated $\Wc$ and $\Zc$ samples, and in $\Wc$ data events. The $\Wc$ distributions are presented after the $\OSSS$ subtraction. Events with $M_\text{vertex}^\text{corr} > 8 \GeV$ are included in the last bin of the corrected secondary-vertex mass distribution.
Vertical bars represent the statistical uncertainties.
}
\label{fig:templates_Wc}
\end{center}
\end{figure*}
The corrected secondary-vertex mass and JP discriminant distributions, normalized to unity, are presented in Fig.~\ref{fig:templates_Wc} for the three analysis categories.
The simulated $\Wc$ and $\Zc$ distributions are compared to $\Wc$ data.
In general, the simulated $\Wc$ and $\Zc$ distributions agree with the $\Wc$ data in all categories.
A noticeable discrepancy is observed between the simulated and measured distributions of the corrected secondary-vertex mass in $\Wc$ events as shown in Fig.~\ref{fig:templates_Wc} (left).
This difference is due to a different fraction of events with two- and three-track vertices in data and in the simulation.
Studies with simulated events demonstrate that the fraction of events with two- and three-track vertices for $\Wc$ and $\Zc$ production is the same.
Therefore, we assume that the $\Wc$ corrected secondary-vertex mass distribution measured in data properly reproduces the same distribution for the $\Zc$ measured events.
The distributions obtained in the electron and muon decay channels are consistent and are averaged to obtain the final templates, thereby decreasing the associated statistical uncertainty.

\begin{figure}[!tb]
\begin{center}
\includegraphics[width=\cmsSingleFigWidth]{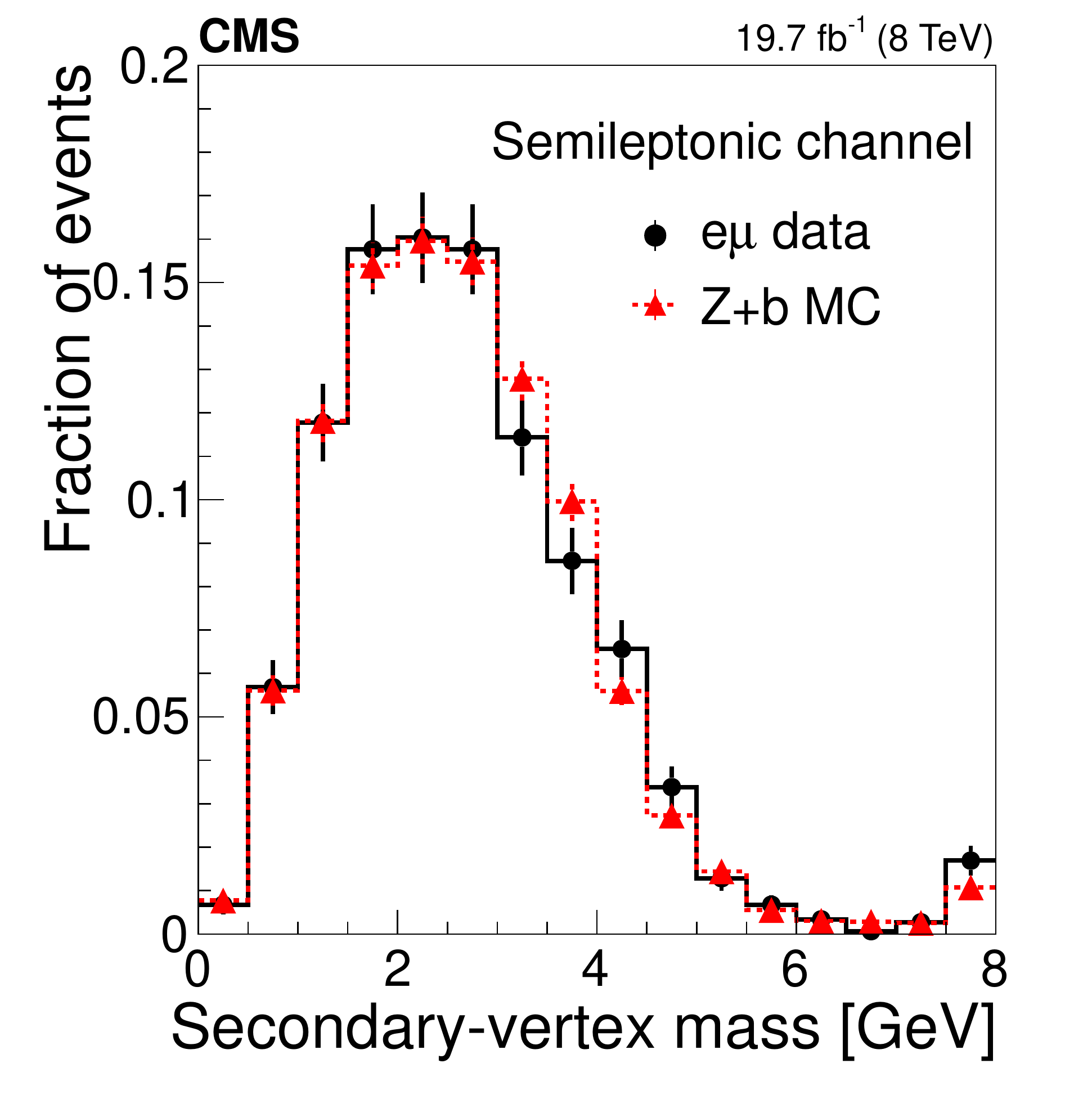}
\caption{Distribution of the corrected secondary-vertex mass normalized to unity from simulated $\Zb$ and $\Pe\PGm$-$\ttbar$ data (described in the text) events.
Vertical bars represent the statistical uncertainties.
The last bin of the distribution includes events with $M_\text{vertex}^\text{corr} > 8\GeV$.
}
\label{fig:vertex_mass_Zb}
\end{center}
\end{figure}

The shape of the discriminant variables for $\Zb$ events is modelled with the simulated samples.
The simulated distribution of the corrected secondary-vertex mass is validated with the sample of $\Pe\PGm$-$\ttbar$ events as shown in
Fig.~\ref{fig:vertex_mass_Zb}.
The simulation describes the data well, apart from the mass regions 3--4 \GeV and above 7.5 \GeV.
The observed differences, ${\approx}13\%$ in the 3--4\GeV mass region and ${\approx}50\%$ above $7.5\GeV$, are used to correct the simulated $\Zb$ distribution.
However, the number of events in the $\Pe\PGm$-$\ttbar$ sample does not allow a validation of the shape of JP discriminant distributions for
$\Zb$ events in the exclusive channels.

{\tolerance=10000
The distributions of the discriminant variables obtained in data are corrected by subtracting the contributions from the various background processes.
They are estimated in the following way:
\begin{itemize}
\item[$\centerdot$] The shapes of the discriminant distributions for $\ttbar$ production are evaluated with the $\Pe\PGm$-$\ttbar$ sample.
The normalization difference between same and different flavour combinations, $N^{\ttbar}_{\Pe\Pe}/N^{\ttbar}_{\Pe\PGm}$ ($N^{\ttbar}_{\PGm\PGm}/N^{\ttbar}_{\Pe\PGm}$) is
estimated from the sideband region $\ETmiss > 80\GeV$, and applied to the signal region $\ETmiss<40\GeV$.
\item[$\centerdot$] The shape and normalization of the corrected secondary-vertex mass distribution for the $\Zuds$ quark background in the semileptonic channel are evaluated with the simulated samples.
Discrepancies between data and simulation in the rate of $\Zuds$ jet misidentification are corrected by applying the appropriate scale factors to the simulation~\cite{CMS-PAS-BTV-13-001}.
No background from the $\Zuds$ quark process is expected in the exclusive channels.
\item[$\centerdot$] The shapes and normalization of the discriminant distributions for the remaining background from diboson production are taken from simulation.
\end{itemize}
\par}

The yields of $\Zc$ and $\Zb$ events in data are estimated by performing least squares fits between the background-subtracted data and template distributions.
Independent fits are performed in the dielectron and dimuon channels and in the three analysis categories.
The expected $\Zc$ and $\Zb$ distributions are fitted to data with scaling factors $\mu_{\Zc}$ and $\mu_{\Zb}$
defined with respect to the initial normalization predicted from simulation as free parameters of the fit.
Typical values of the scaling factors are in the range 0.95--1.05 with a correlation coefficient between $\mu_{\Zc}$ and $\mu_{\Zb}$
of the order of $-0.4$. The scaling factor obtained for the $\Zb$ component
is consistent with that reported in Ref.~\cite{Khachatryan:2016iob} for a similar fiducial region.
The fitted $\mu_{\Zc}$ and $\mu_{\Zb}$ are applied to the expected yields to obtain the measured ones in the data.
The measured yields are summarized in Table~\ref{table:Cross_sections}.

\begin{table*}[tbp]
\begin{center}
\topcaption{Cross section $\SZc \,\mathcal{B}$ and cross section ratio $\SZc/\SZb$ in the three categories of this analysis and in the two $\PZ$ boson decay channels. The $N^{\rm signal}_{\Zc}$ and $N^{\rm signal}_{\Zb}$ are the yields of $\Zc$ and $\Zb$ events, respectively, extracted from the fit to the corrected secondary-vertex mass (semileptonic mode) or JP discriminant ($\Dpm$ and $\Dstar$ modes) distributions.
The factors $\mathcal{C}$ that correct the selection inefficiencies are also given. They include the relevant branching fraction for the corresponding channel.
All uncertainties quoted in the table are statistical, except for those of the measured cross sections and cross section ratios where the first uncertainty is statistical and the second is the estimated systematic uncertainty from the sources discussed in the text.}
\renewcommand{\arraystretch}{1.2}
\begin{tabular}{c|c|c|c}
\multicolumn{4}{c}{Semileptonic mode} \\
\hline
Channel & $N^{\rm signal}_{\Zc}$ & $\mathcal{C}_{\Zc}$ (\%) & $\SZc\,\mathcal{B}$ [\unit{pb}] \\
\hline
$\PZ \to {\Pe^+}{\Pe^-}$ & 1070 $\pm$ 100 & $0.63\pm 0.03$ & 8.6 $\pm$ 0.8 $\pm$ 1.0 \\
$\PZ \to {\PGm^+}{\PGm^-}$ & 1450 $\pm$ 140 & $0.81\pm 0.03$ & 9.1 $\pm$ 0.9 $\pm$ 1.0 \\
\hline
\rule{0pt}{1.2em} $\PZ \to {\ell^+}{\ell^-}$ & \multicolumn{3}{c}{$\SZc\,\mathcal{B} =  8.8 \pm 0.6 \stat \pm 1.0 \syst \unit{pb}$} \\ [3pt]
\hline
\hline
Channel & $N^{\rm signal}_{\Zb}$ & $\mathcal{C}_{\Zb}$ (\%) & $\SZc/\SZb$ \\
\hline
$\PZ \to {\Pe^+}{\Pe^-}$ & 2610 $\pm$ 110 & $2.90\pm 0.08$ & 1.9 $\pm$ 0.2 $\pm$ 0.2\\
$\PZ \to {\PGm^+}{\PGm^-}$ & 3240 $\pm$ 150 & $3.93\pm 0.10$ & 2.2 $\pm$ 0.3 $\pm$ 0.2\\
\hline
\rule{0pt}{1.2em} $\PZ \to {\ell^+}{\ell^-}$ & \multicolumn{3}{c}{$\SZc/\SZb = 2.0 \pm 0.2 \stat \pm 0.2 \syst$} \\ [3pt]
\hline
\hline
\multicolumn{4}{c}{$\Dpm$ mode} \\
\hline
Channel & $N^{\rm signal}_{\Zc}$ & $\mathcal{C}_{\Zc}$ (\%) & $\SZc\,\mathcal{B}$ [pb] \\
\hline
$\PZ \to {\Pe^+}{\Pe^-}$ & 280 $\pm$ 60 & $0.13\pm 0.02$ & 10.9 $\pm$ 2.2 $\pm$ 0.9 \\
$\PZ \to {\PGm^+}{\PGm^-}$ & 320 $\pm$ 80 & $0.18\pm 0.02$ &  8.8 $\pm$ 2.0 $\pm$ 0.8 \\
\hline
\rule{0pt}{1.2em} $\PZ \to {\ell^+}{\ell^-}$ & \multicolumn{3}{c}{$\SZc\,\mathcal{B} = 9.7 \pm 1.5 \stat \pm 0.8 \syst \unit{pb}$} \\ [3pt]
\hline
\hline
\multicolumn{4}{c}{$\Dstar$ mode} \\
\hline
Channel & $N^{\rm signal}_{\Zc}$ & $\mathcal{C}_{\Zc}$ (\%) & $\SZc\,\mathcal{B}$ [\unit{pb}] \\
\hline
$\PZ \to {\Pe^+}{\Pe^-}$ & 150 $\pm$ 30 & $0.11\pm 0.01$ & 7.3 $\pm$ 1.5 $\pm$ 0.5 \\
$\PZ \to {\PGm^+}{\PGm^-}$ & 250 $\pm$ 30 & $0.14\pm 0.01$ & 9.3 $\pm$ 1.1 $\pm$ 0.7 \\
\hline
\rule{0pt}{1.2em} $\PZ \to {\ell^+}{\ell^-}$ & \multicolumn{3}{c}{$\SZc\,\mathcal{B} = 8.5 \pm 0.9 \stat \pm 0.6 \syst \unit{pb}$} \\ [3pt]
\hline
\hline
\multicolumn{4}{c}{Combination} \\
\hline
\rule{0pt}{1.2em} $\PZ \to {\ell^+}{\ell^-}$ & \multicolumn{3}{c}{$\SZc\,\mathcal{B} = 8.8 \pm 0.5 \stat \pm 0.6 \syst \unit{pb}$} \\ [3pt]
\end{tabular}
\label{table:Cross_sections}
\end{center}
\end{table*}

\begin{figure*}[!tb]
\begin{center}
\includegraphics[width=\cmsDoubleFigWidth]{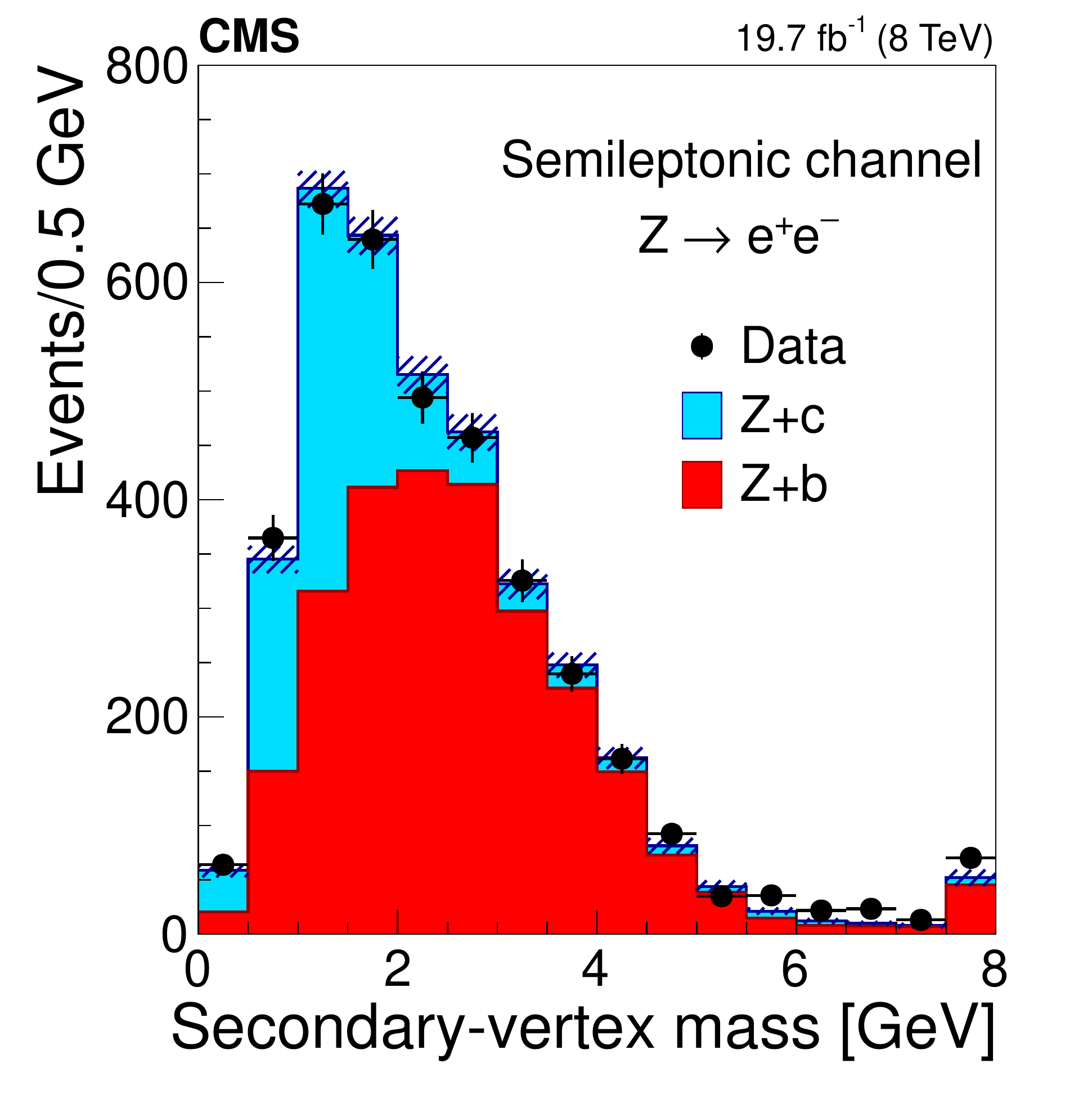}
\includegraphics[width=\cmsDoubleFigWidth ]{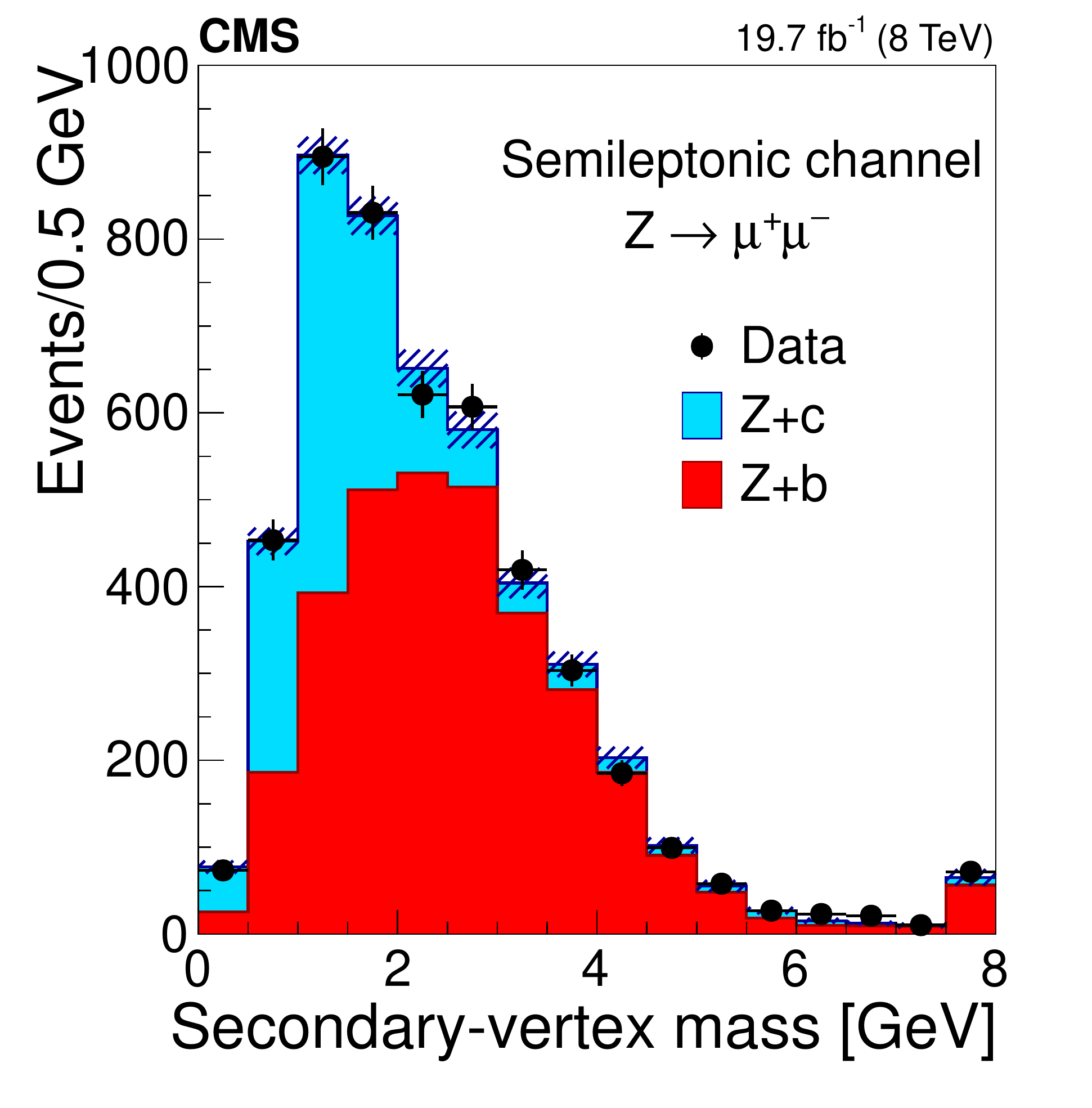}
\caption{Corrected secondary-vertex mass distributions, after background subtraction,
in the dielectron (left) and dimuon (right) channels for events selected in the semileptonic mode.
Events with $M_\text{vertex}^\text{corr}>8 \GeV$ are included in the last bin of the distribution.
The shape of the $\Zc$ and $\Zb$ contributions is estimated as explained in the text.
Their normalization is adjusted to the result of the signal extraction fit.
Vertical bars on data points represent the statistical uncertainty in the data. The hatched areas represent the sum in quadrature of the statistical uncertainties of the templates
describing the two contributions ($\Zc$ from $\Wc$ data events and $\Zb$ from simulation).
}
\label{fig:vertex_mass}
\end{center}
\end{figure*}

Figure~\ref{fig:vertex_mass} shows the background-subtracted distributions of the corrected
secondary-vertex mass for the $\Zj$ events with a \textit{muon-inside-a-jet} associated with a secondary vertex.
The corrected secondary-vertex mass tends to be larger for $\Zb$ than for $\Zc$ events because the larger
mass of the $\PQb$ quark gives rise to heavier hadrons ($m_{\PQb\,\text{hadrons}} \approx 5\GeV$, $m_{\PQc\,\text{hadrons}}  \approx 2\GeV$).

\begin{figure*}[!tb]
\begin{center}
\includegraphics[width=\cmsDoubleFigWidth]{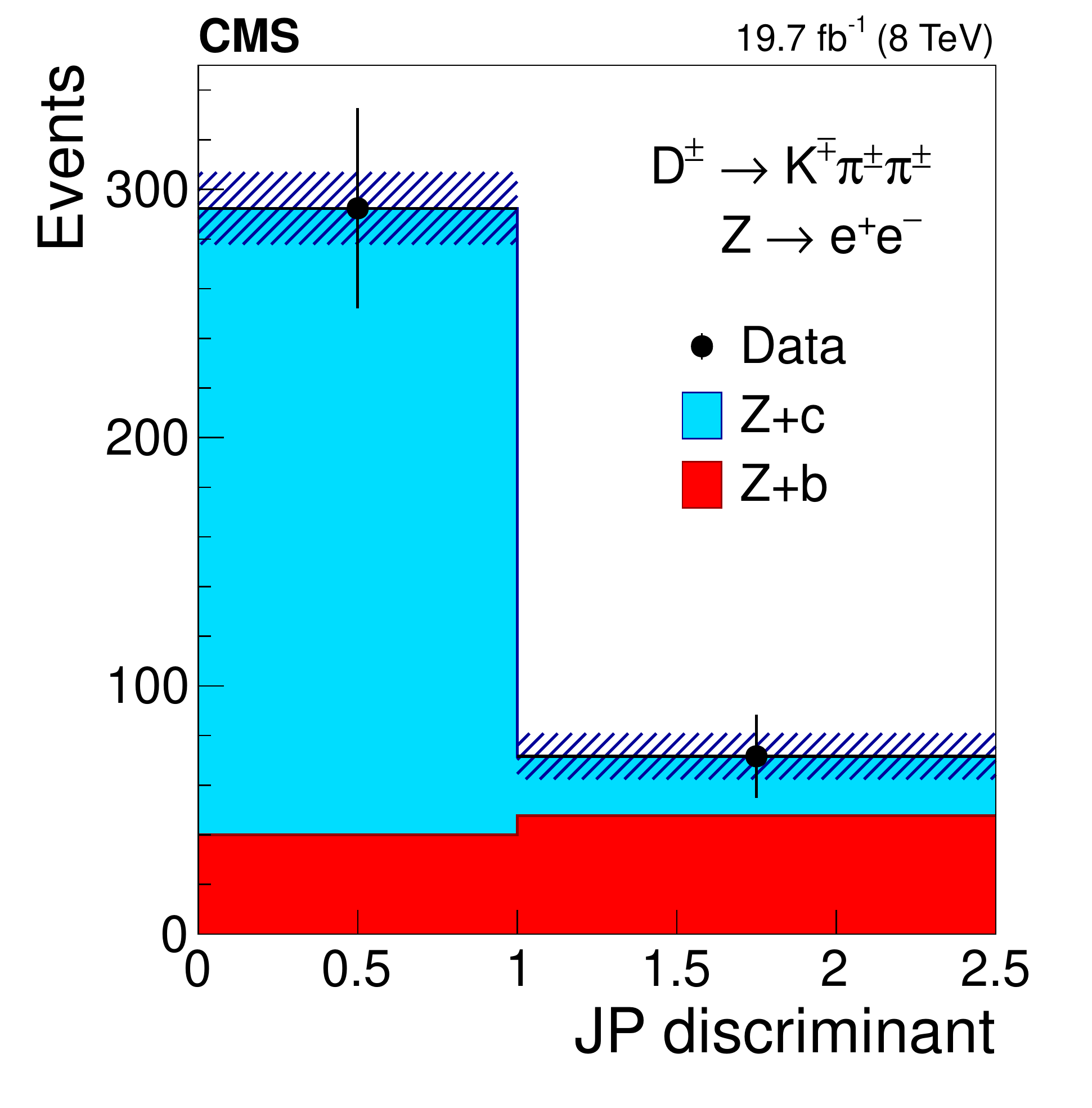}
\includegraphics[width=\cmsDoubleFigWidth]{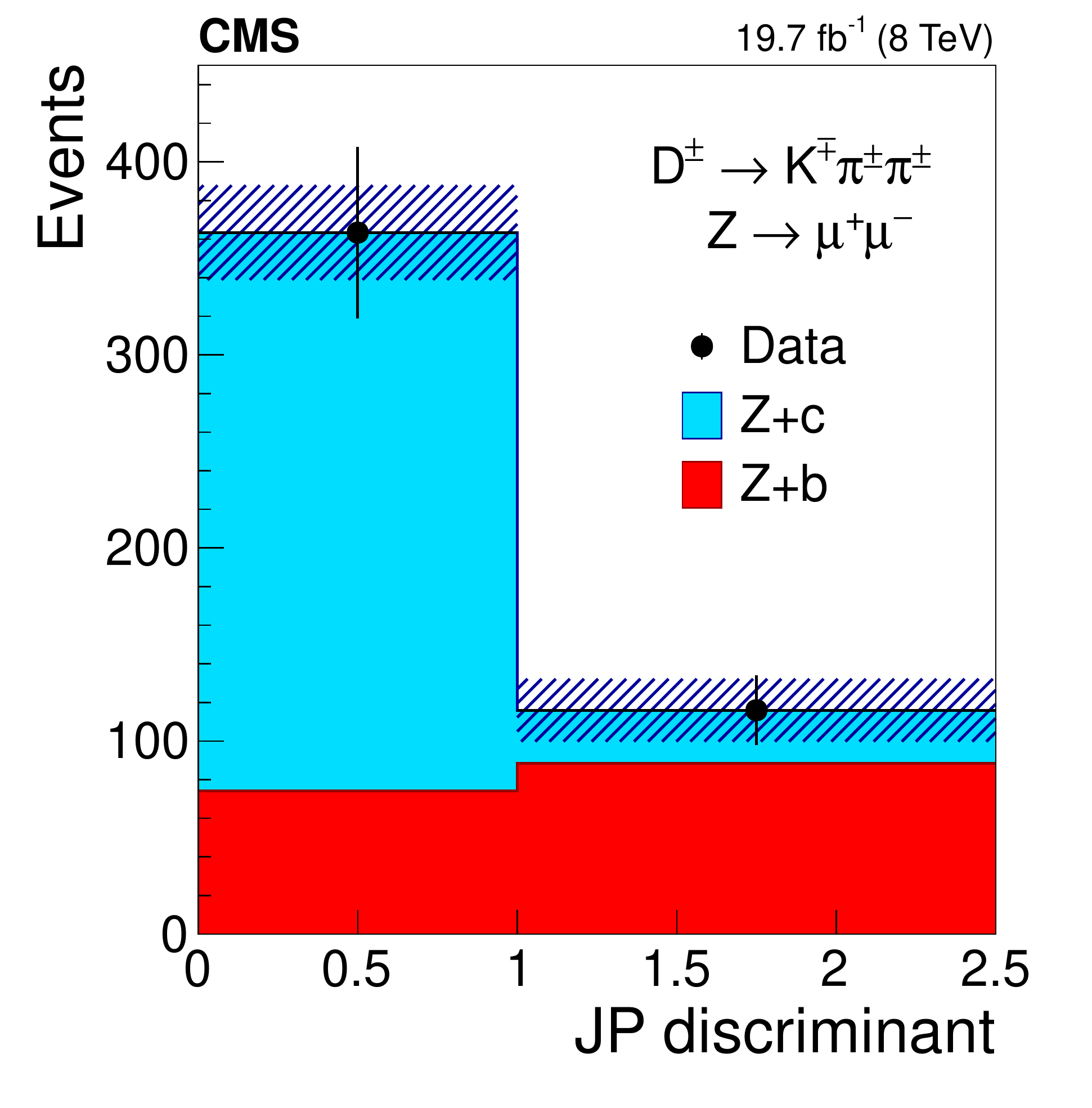}
\caption{Background-subtracted distributions of the JP discriminant in the dielectron (left) and dimuon (right) channels
for $\Zj$ events with a $\Dpm \to \PK^\mp \PGp^\pm\PGp^\pm$ candidate.
The shape of the $\Zc$ and $\Zb$ contributions is estimated as explained in the text.
Their normalization is adjusted to the result of the signal extraction fit.
Vertical bars on data points represent the statistical uncertainty in the data. The hatched areas represent the sum in quadrature of the statistical uncertainties of the templates
describing the two contributions ($\Zc$ from $\Wc$ data events and $\Zb$ from simulation).
}
\label{fig:JPDPM}
\end{center}
\end{figure*}

The JP discriminant takes lower values for $\Zc$ events than for $\Zb$ events. The $\Dpm$ or $\Dstar$ mesons in $\Zb$ events are
``secondary'' particles, \ie they do not originate from the hadronization of a $\PQc$ quark produced at the primary vertex,
but are decay products of previous $\bhadron$ decays at unobserved secondary vertices.
Figure~\ref{fig:JPDPM} shows the background-subtracted distribution of the JP discriminant for the $\Zj$ events with a $\Dpm \to \PK^\mp \PGp^\pm\PGp^\pm$ candidate.
Two bins are used to model the JP discriminant in this channel; as a result, the determination of the scaling factors $\mu_{\Zc}$ and $\mu_{\Zb}$ is reduced to solving
a system of two equations with two unknowns.

\begin{figure*}[!tb]
\begin{center}
\includegraphics[width=\cmsDoubleFigWidth]{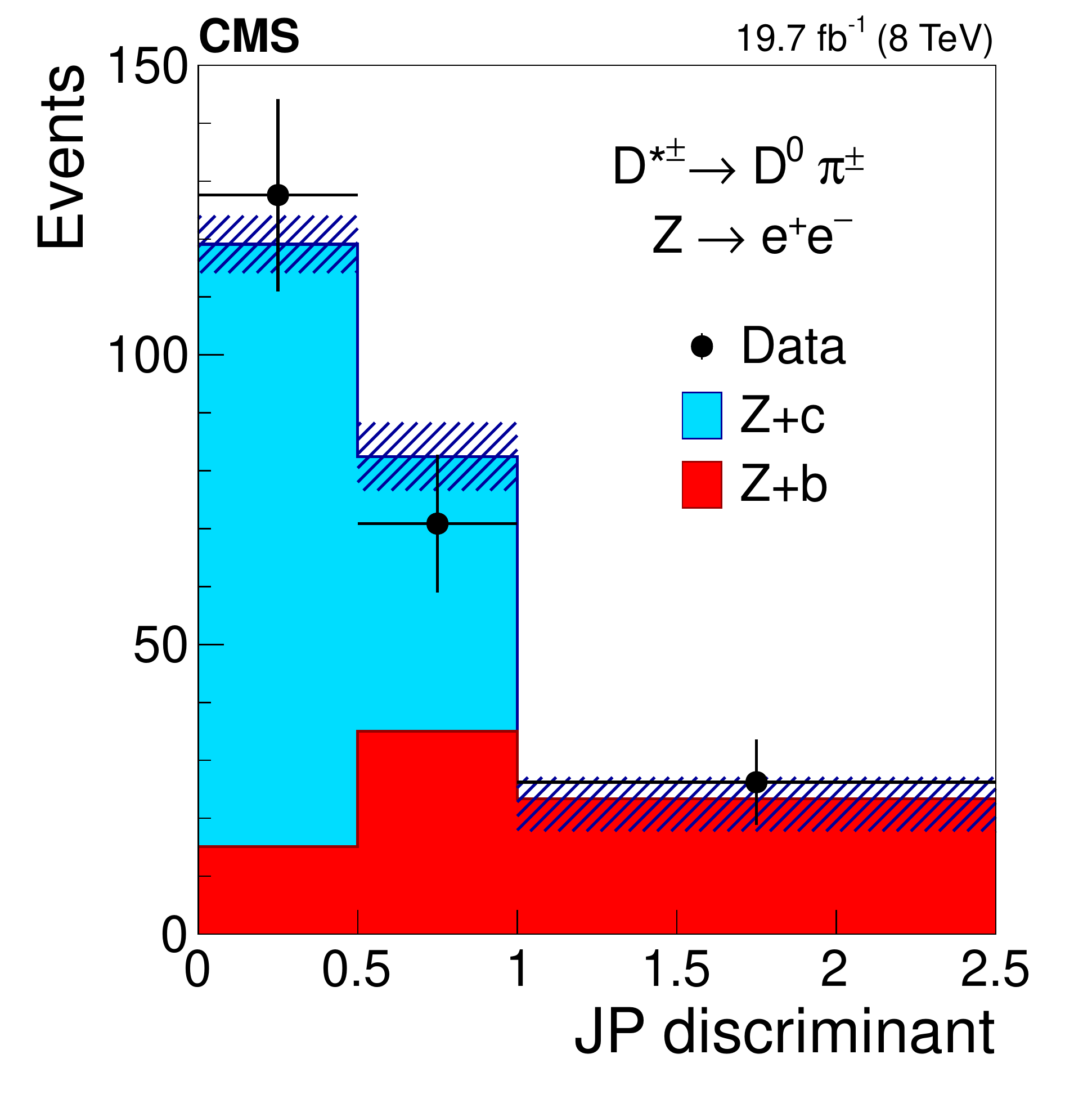}
\includegraphics[width=\cmsDoubleFigWidth]{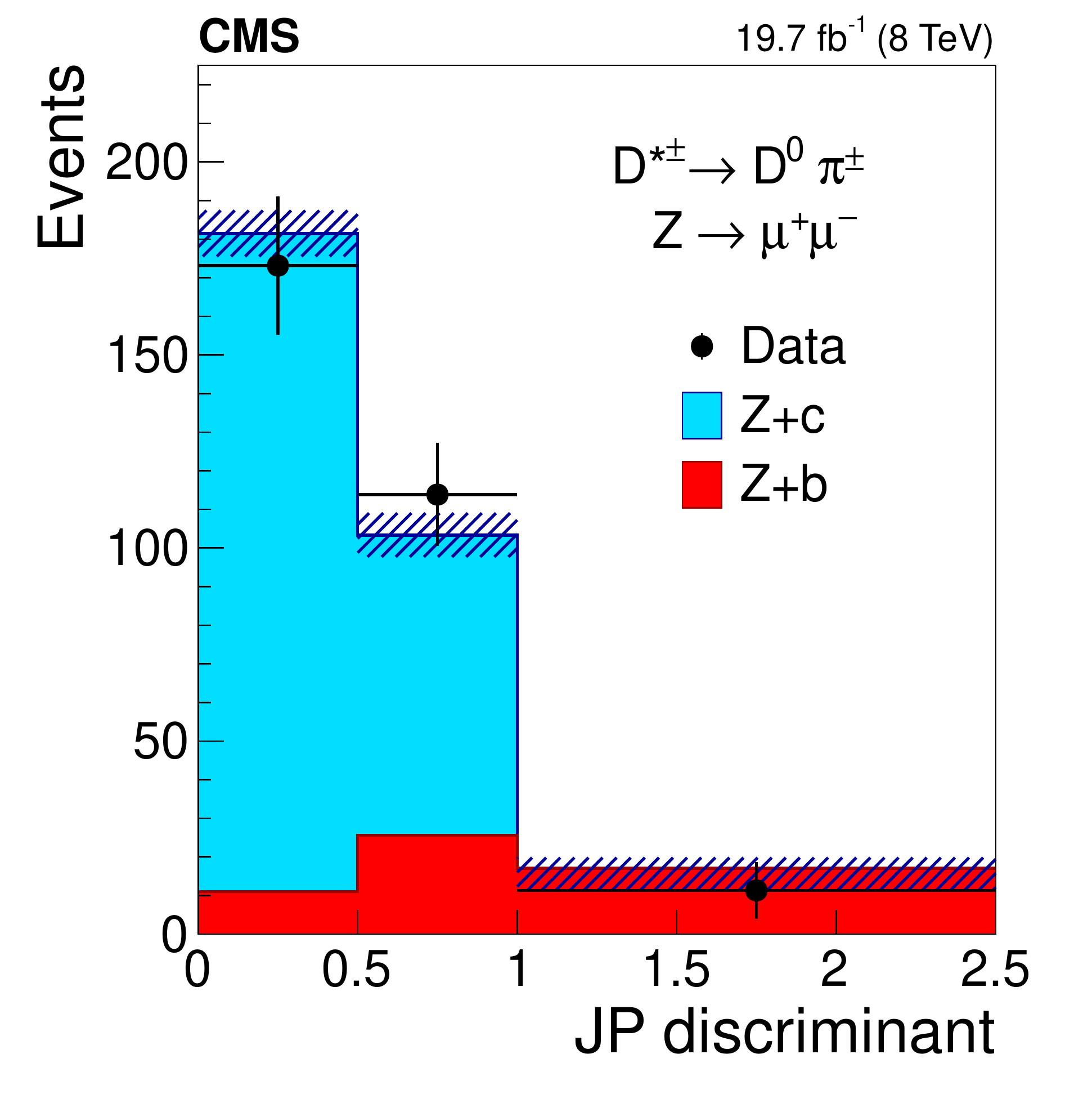}
\caption{Background-subtracted distributions of the JP discriminant in the dielectron (left) and dimuon (right) channels
for $\Zj$ events with a $\Dstar \to\Dz\PGp^\pm \to \PK^\mp \PGp^\pm\PGp^\pm$ candidate.
The shape of the $\Zc$ and $\Zb$ contributions is estimated as explained in the text.
Their normalization is adjusted to the result of the signal extraction fit.
Vertical bars on data points represent the statistical uncertainty in the data. The hatched areas represent the sum in quadrature of the statistical uncertainties of the templates
describing the two contributions ($\Zc$ from $\Wc$ data events and $\Zb$ from simulation).
}
\label{fig:JPDSTAR}
\end{center}
\end{figure*}

Figure~\ref{fig:JPDSTAR} presents the background-subtracted distribution of the JP discriminant for the $\Zj$ events with a $\Dstar$ candidate.
In this latter channel the particle identified as the \textit{soft pion} in the $\Dstar\to\Dz\PGp^\pm$ decay is a true primary particle
in the case of $\Zc$ events, whereas it arises from a secondary decay ($\bhadron \to \Dstar+\PX \to\Dz\PGp^\pm+\PX$) for $\Zb$ events.
This ``secondary'' origin of the \textit{soft pion} generates a distinctive dip in the first bin of the JP discriminant distribution for $\Zb$ events.

\section{Systematic uncertainties~\label{sec:syst_uncert}}

\begin{figure}[!tb]
\begin{center}
\includegraphics[width=\cmsSingleFigWidth]{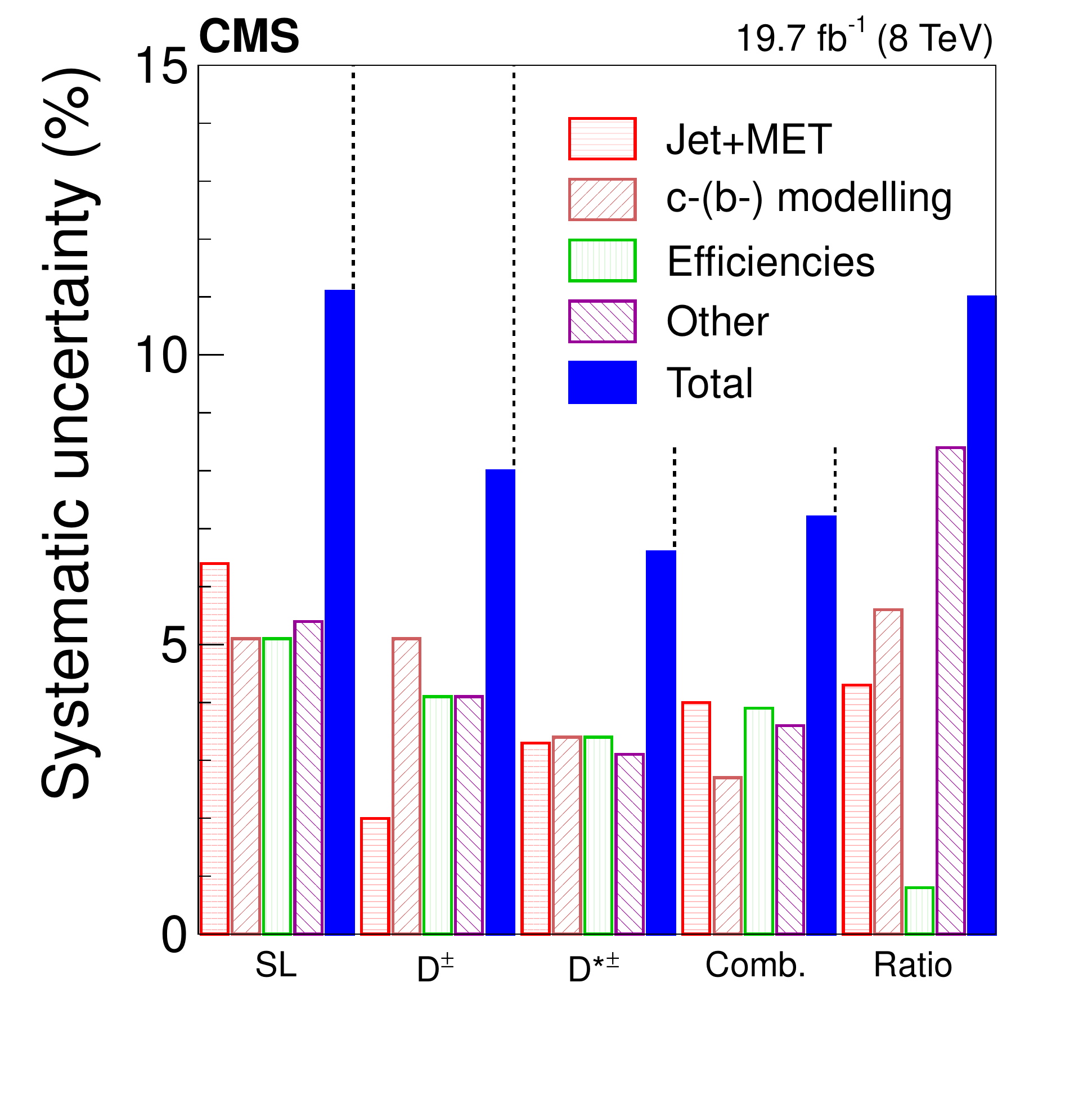}
\caption{Contributions to the systematic uncertainty in the measured $\Zc$ cross section and in the $\ZcZb$ cross section ratio.
The first three blocks in the graph show the uncertainties in the $\Zc$ cross section in the three decay modes, semileptonic (SL), $\Dpm$, and $\Dstar$,
calculated from the combination of the dimuon and dielectron $\PZ$ boson decay channels.
The fourth block shows the systematic uncertainties in the combined (Comb.) $\Zc$ cross section. The last block
presents the systematic uncertainty in the $\ZcZb$ cross section ratio measured in the semileptonic mode.
For every block, the height of the hatched bars indicates the contribution from the different sources of systematic uncertainty. The last, solid bar
shows their sum in quadrature.}
\label{figsys}
\end{center}
\end{figure}

Several sources of systematic uncertainties are identified, and their impact on the measurements is estimated
by performing the signal extraction fit with the relevant parameters in the simulation varied up and down by their uncertainties.
The effects are summarized in Fig.~\ref{figsys}.
The contributions from the various sources are combined into fewer categories for presentation in Fig.~\ref{figsys}.

One of the main uncertainties is related to the charm fractions for the production
and decay of $\chadrons$ in the simulated samples and to the determination of the $\PQc$ tagging efficiency.
The average of the inclusive charm quark semileptonic branching fractions is ${\mathcal{B}}(\PQc\to\ell) = 0.096\pm 0.004$~\cite{PDG},
and the exclusive sum of the individual contributions from all weakly decaying charm hadrons is $0.086\pm 0.004$~\cite{gladilin,PDG}.
The average of these two values, ${\mathcal{B}}(\PQc\to\ell) = 0.091 \pm 0.003$, is consistent with the \PYTHIA value
used in our simulations (9.3\%).
We assign a 5\% uncertainty in order to cover both central values within one standard deviation.
The average of the inclusive $\PQb$ quark semileptonic branching fractions is ${\mathcal{B}}(\PQb\to\ell) = 0.1069 \pm 0.0022$~\cite{PDG},
which is consistent with the \PYTHIA value used in our simulations (10.5\%).
The corresponding uncertainty of 2\% is propagated. The 5\% systematic uncertainty in ${\mathcal{B}}(\PQc\to\ell)$ is further propagated for the
fraction of $\Zb$ events with a lepton in the final state through the decay chain $\PQb\to\PQc\to\ell$.
Uncertainties in the branching ratios of other $\PQb$ hadron decay modes with a lepton in the final state,
such as $\bhadron\to\tau(\hspace{-.12em}\to\hspace{-.14em}\ell\hspace{-.12em}+\hspace{-.14em}\PX)\hspace{-.12em}+\hspace{-.14em}\PX^\prime$,
$\bhadron\to{\mathrm{J}\hspace{-.08em}/\hspace{-.14em}\psi}(\hspace{-.12em}\to\hspace{-.14em} \ell^+\ell^-)\hspace{-.12em}+\hspace{-.14em}\PX$,
are not included since the expected contribution to the selected sample is negligible.

Since the simulation in the $\Dpm$ and $\Dstar$ modes is reweighted to match the experimental values~\cite{gladilin},
the uncertainty in the reweighting factors (5\% for $\Dpm$ and 3.2\% for $\Dstar$) is propagated to the cross section.

The contribution from gluon splitting processes to $\Zc$ production in the phase space of the measurement
is small, and its possible mismodelling has little impact on the measurements.
Its effect is evaluated with the simulated sample by independently increasing the weight of the events with at least two $\PQc$ ($\PQb$) quarks in the list
of generated particles close to the selected jet ($\Delta R ({\text{jet}},\PQc (\PQb)) < 0.5$)
by three times the experimental uncertainty in the gluon splitting rate into $\ccbar$, $\bbbar$ quark pairs~\cite{Alephgccbar, Alephgbbbar}.

The effects of the uncertainty in the jet energy scale and jet energy resolution are assessed by varying the corresponding jet energy scale
(jet energy resolution) correction factors within their uncertainties according to the results of dedicated CMS studies~\cite{CMS-PAPER-JME-10-011, CMS-PAPER-JME-13-004}.
The uncertainty from a mismeasurement of the missing transverse energy in the event is estimated by propagating the jet energy scale uncertainties and by adding 10\% of the energy unassociated with reconstructed PF objects to the reconstructed $\ETmiss$.

The uncertainty in the $\PQc$ tagging scale factors is in the range 3.5--4\%, and it is around 2.5\% for the $\PQb$ tagging efficiency.
In the $\Dstar$ mode, the candidate reconstruction procedure is repeated by independently changing by one standard deviation, in terms of the $\pt$ resolution,
the different $\pt$-thresholds imposed and the decay length significance requirement.
We assume the uncertainty is the quadratic sum of the respective differences between data and simulation in the change of the number of $\Dstar$ candidates (2.8\%).

The uncertainty in the lepton efficiency correction factors is 4\% in the $\Zee$ and 2\% in the $\Zmm$ channels.
The uncertainty in the efficiency for the identification of muons inside jets is approximately 3\%, according to dedicated studies in multijet events~\cite{CMS-PAPER-MUO-10-004}.

An additional systematic uncertainty is assigned to account for a possible mismodelling of the subtracted backgrounds.
For the $\ttbar$ background the uncertainty is taken as the difference between the estimate based on data, as described in Section~\ref{sec:sigext}, and the one based on simulation.
For $\Zuds$ events, the systematic uncertainty is evaluated by using the MC correction factors associated with different misidentification probabilities.
Finally, the diboson contribution is varied by the difference between the theoretical cross sections calculated at NNLO and NLO ($\approx$15\%)~\cite{Gehrmann:2014fva, Cascioli:2014yka, Grazzini:2016swo}.

The reference signal simulated sample is generated with \MADGRAPH+\PYTHIAsix using the PDF CTEQ6L1 and reweighted to NNLO PDF set MSTW2008NNLO.
The difference resulting from using other NNLO PDF sets is small (${\lesssim}1\%$).
Following the prescription of the PDF groups, the PDF uncertainty is of the same order.

The shapes of the discriminant distributions obtained from the $\Wc$ event sample are observed to be very stable.
Changes in the jet energy scale and variations in the $\pt$ threshold imposed to select $\PW$ boson candidates do not affect the shape of the templates.
The correction factors applied in certain regions to the corrected secondary-vertex mass template for $\Zb$ events are varied within their uncertainties.

Uncertainties due to the pileup modelling are calculated using a modified pileup profile obtained with a $\pp$ inelastic cross section changed by its estimated uncertainty, 6\%.
The uncertainty in the determination of the integrated luminosity of the data sample is 2.6\%~\cite{CMS-PAS-LUM-13-001}.

Systematic uncertainties in the differential $\Zc$ cross section and in the $\ZcZb$ cross section ratio are in the range 11--15\%.
The main sources of systematic uncertainty in the differential distributions are due to the jet energy scale determination, the charm fractions for $\chadron$
production and decay in simulation, and the efficiencies of heavy flavour tagging.
The uncertainty in the binned $\PQc$ tagging efficiency scaling factors is 7--8\%. Uncertainties in the $\PQb$ tagging efficiencies are 3--5\%.
An additional source of systematic uncertainty in the differential measurement as a function of the transverse momentum of the jet arises from the statistical
uncertainty in the determination of the response matrix used to correct for migration of events across $\pt^{\jet}$ bins, as described in Section~\ref{sec:xsec_diff}.
Its impact is evaluated by repeating the correction procedure using a large number of response matrices, built from the nominal one by varying its components according to their statistical uncertainties. The effect is in the range 4--6\% for the $\Zc$ cross section and 4.5--7\% for the $\ZcZb$ cross section ratio.

\section{Inclusive \texorpdfstring{$\Zc$}{Z + c} cross section and \texorpdfstring{$\ZcZb$}{(Z + c)/(Z + b)} cross section ratio~\label{sec:xsec_inc}}

For all channels under study, the $\Zc$ cross section is determined in the fiducial region $\pt^{\ell} > 20\GeV$, $|\eta^{\ell}|<2.1$,
$71 < m_{\ell\ell} < 111\GeV$, $\pt^{\jet}>25\GeV$, $|\eta^{\jet}|<2.5$, and $\Delta R ({\text{jet}},\ell) > 0.5$, using the following expression:
\begin{equation}
\SZc\,\mathcal{B} = \frac{N^{\rm signal}_{\Zc}}{\mathcal{C} \, \mathcal{L}},\label{total_sigma}
\end{equation}
where $N^\text{signal}_{\Zc}$ is the fitted yield of $\Zc$ events and $\mathcal{L}$ is the integrated luminosity.
The factor $\mathcal{C}$ corrects for event losses in the selection process and is estimated using simulated events.
The $\mathcal{C}$  factors also include the relevant branching fraction for the corresponding channel.

{\tolerance = 1200
Similarly, the ratio of cross sections $\sigma(\Zc)/\sigma(\Zb)$ is calculated in the same fiducial region applying the previous expression also for the $\Zb$ contribution:
\begin{equation}
\frac{\SZc}{\SZb} = \frac{N^{\rm signal}_{\Zc}}{N^{\rm signal}_{\Zb}} \, \frac{\mathcal{C}(\Zb)}{\mathcal{C}(\Zc)},\label{ratio_sigma}
\end{equation}
Table~\ref{table:Cross_sections} shows the $\Zc$ production cross section obtained in the three modes
and the $\ZcZb$ cross section ratio (semileptonic mode only).
\par}

For the three categories of this analysis the $\Zc$ cross sections obtained in the dielectron and dimuon $\PZ$ boson decay channels are consistent.
The results obtained in the three analysis categories are also consistent. Several combinations are performed to improve the precision of the
measurement taking into account statistical and systematic uncertainties of the individual measurements.
Systematic uncertainties arising from a common source and affecting several measurements are considered as fully correlated.
In particular, all systematic uncertainties are assumed fully correlated between the electron and muon channels, except those related to lepton reconstruction.
The average $\Zc$ cross sections obtained in the three categories, together with the combination of the six measurements, are also presented in Table~\ref{table:Cross_sections}.
The combination is dominated by the result in the semileptonic mode. The contribution of the $\Dstar$ mode to the average is also significant despite the
limited size of the selected samples.

The cross section ratio $\SZc/\SZb$
has been measured in the semileptonic mode, in the two $\PZ$ boson decay channels, and the results among them are consistent.
Both cross section ratios are combined taking into account the statistical and systematic uncertainties in the two channels, and the correlations among them.
The combination is given in Table~\ref{table:Cross_sections}.

The measured $\Zc$ cross section and the $\ZcZb$ cross section ratio are compared to theoretical predictions obtained using two MC
event generators and the \MCFM program.

A prediction of the $\Zc$ fiducial cross section is obtained with the \MADGRAPH sample.
It is estimated by applying the phase space definition requirements to generator level quantities:
two leptons from the $\PZ$ boson decay with $\pt^{\ell}>20\GeV$, $|\eta^{\ell}| < 2.1$, and
dilepton invariant mass in the range $71 < m_{\ell\ell} < 111\GeV$;
a generator-level $\cjet$ with $\pt^{\cjet} > 25\GeV$, $|\eta^{\cjet}| < 2.5$
and separated from the leptons by a distance $\Delta R (\cjet,\ell) > 0.5$.
A prediction of the $\Zb$ cross section, and hence of the $\ZcZb$ cross section ratio, is similarly derived
applying the relevant phase space definition requirements to $\PQb$ flavoured generator-level jets.

{\tolerance=1000
The \MADGRAPH prediction, $\SZc\,\mathcal{B}
= 8.14\pm 0.03~\stat \pm 0.25~({\rm PDF}) \unit{pb}$, is in agreement with the measured value.
The quoted PDF uncertainty corresponds to the largest difference in the predictions obtained
using the central members of two different PDF sets (MSTW2008 vs NNPDF2.3);
uncertainties computed using their respective PDF error sets are about half this value.
\par}

We have also compared the measurements with predictions obtained with a sample of events generated with \MGvATNLO v2.2.1~\cite{Alwall:2014hca}
(hereafter denoted as \aMCatNLO) generator interfaced with \PYTHIA v8.212~\cite{Pythia8} using the CUETP8M1
tune~\cite{Khachatryan:2015pea} for parton showering and hadronization.
The matrix element calculation includes the $\PZ$ boson production process with 0, 1, and 2 partons at NLO.
The FxFx~\cite{Frederix:2012ps} merging scheme between jets from matrix element and parton showers is implemented with a merging scale parameter set to 20\GeV.
The NNPDF3.0 PDF set~\cite{Ball:2014uwa} is used for the matrix element calculation, while the NNPDF2.3 LO is used for the
showering and hadronization.

The \aMCatNLO prediction of the $\Zc$ cross section is slightly higher,
$\SZc\,\mathcal{B}
= 9.46\pm 0.04\stat \pm 0.15\,(\text{PDF}) \pm 0.50\,(\text{scales})\unit{pb}$,
but still in agreement with the measurement.
Uncertainties in the prediction are evaluated using the reweighting features implemented in the generator~\cite{Frederix:2011ss}.
The quoted PDF uncertainty corresponds to the standard deviation of the predictions obtained using the one hundred replicas in the NNPDF3.0 PDF set.
The scale uncertainty is the envelope of the predictions when the factorization and renormalization scales are varied by a factor
of two or one half independently, always keeping the ratio between them less than or equal to two.

Theoretical predictions in perturbative quantum chromodynamics at NLO for the associated production of a $\PZ$ boson and at least one $\PQc$ quark are obtained
with the \MCFM~7.0 program~\cite{Campbell:2003dd}. Several sets of NLO PDF sets are used, accessed through the LHAPDF6~\cite{Buckley:2014ana} library interface.
Partons are clustered into jets using the anti-$\kt$ algorithm with a distance parameter of 0.5.
The kinematic requirements follow the experimental selection: the two leptons from the $\PZ$ boson decay with $\pt^\ell > 20\GeV$, $\abs{\eta^\ell} < 2.1$, $71 < m_{\ell\ell} < 111\GeV$
and a $\PQc$ parton jet with $\pt^{\, \parton\,\jet} > 25\GeV$, $\abs{\eta^{\parton\,\jet}} < 2.5$, and separated from the leptons by $\Delta R (\parton\,\jet,\ell) > 0.5$.
The factorization and renormalization scales are set to the mass of the $\PZ$ boson.
The PDF uncertainty in the predictions is evaluated following the prescription recommended by the individual PDF groups;
the scale uncertainty is estimated as the envelope of the results with (twice, half) factorization and renormalization scales variations.

The prediction computed with \MCFM follows the calculation reported in Refs.~\cite{Campbell:2003dd,Campbell:2002zm}. The leading contribution $\gcLO$
is evaluated at NLO including virtual and real corrections. Some of these corrections feature two jets in the final state, one of them with
heavy flavour quark content.
The calculation also includes the process $\qqZcc$ evaluated at LO, where either one of the heavy flavour quarks escapes detection or the two of them coalesce into a single jet.

The \MCFM prediction, which is a parton-level calculation, is corrected for hadronization effects so it can be compared with the particle-level measurements
reported in this paper.
The correction factor is computed with the \MADGRAPH simulated sample comparing the predicted cross section using generator-level jets
and parton jets. Parton jets are defined using the same anti-$\kt$ clustering algorithm with a distance parameter of 0.5,
applied to all quarks and gluons after showering, but before hadronization.
The flavour assignment for parton jets follows similar criteria as for generator-level jets:
a parton jet is labelled as a b jet if there is at least a b quark among its constituents, regardless of the presence of any c or light quarks.
It is classified as c jet if there is at least a c quark, and no b quark, among the constituents, and light otherwise.
The size of the correction is ${\approx}10\%$ for $\Zc$ and ${\approx}15\%$ for $\Zb$ cross sections, in good agreement with the estimation in Ref.~\cite{Chatrchyan:2012vr}.

{\tolerance=1200
After the hadronization correction the \MCFM prediction still misses contributions from the parton show\-er evolution, underlying event,
and multiple parton interactions.
An approximate value of the total correction due to these processes and hadronization is estimated using \MADGRAPH and amounts to $\approx 30\%$.
This correction is not applied to \MCFM predictions, but can explain the observed differences between \MCFM and the predictions of other generators.
\par}

Predictions are produced using MSTW08 and CT10 PDF sets and a recent PDF set from the NNPDF Collaboration, NNPDF3IC~\cite{Ball:2016neh},
where the charm quark PDF is no longer assumed to be perturbatively generated through pair production from gluons and light quarks,
but is parameterized and determined along with the light quark and gluon PDFs. The PDF set where the charm quark PDF is generated perturbatively,
NNPDF3nIC~\cite{Ball:2016neh}, is also used.

No differences in the predictions are observed using either NNPDF3IC or NNPDF3nIC PDF sets.
Differences among them start to be sizeable when the transverse momentum of the $\PZ$ boson is ${\gtrsim}100\GeV$~\cite{Ball:2016neh}.
The largest prediction is obtained using the MSTW08 PDF set,
$\SZc\,\mathcal{B}
= 5.32 \pm 0.01 \stat~^{+0.12}_{-0.06}\,(\mathrm{PDF})~^{+0.34}_{-0.38}\,(\text{scales})\unit{pb}$.
Predictions obtained using CT10 and NNPDF3IC are 5\% smaller than with MSTW08. The uncertainties in all the calculations are of the same order.

The \MADGRAPH prediction for the $\ZcZb$ cross section ratio is $1.781 \pm 0.006 \stat \pm 0.004~({\rm PDF})$,
where the PDF uncertainty reflects the largest variation using the various PDF sets.
The expectation from \aMCatNLO is $1.84 \pm 0.01 \stat \pm 0.07~({\rm scales})$.
The uncertainties from the several members within one PDF set essentially vanish in the ratio.
Both predictions agree with the measured ratio.

A prediction for the cross section ratio is also obtained with \MCFM, as the ratio of the predictions for $\SZc$ and $\SZb$, using the same
parameters emulating the experimental scenario for both processes. The calculation of the $\SZb$ cross section follows the same reference as
$\SZc$~\cite{Campbell:2003dd,Campbell:2002zm}.
The highest predicted value is
$\SZc/\SZb
= 1.58 \pm 0.01\,(\text{stat+PDF syst}) \pm 0.07\,(\text{scales})$
obtained when the CT10 PDF set is used.
The prediction from NNPDF3IC is about 10\% lower, mainly because the predicted $\Zb$ cross section using this PDF is the highest one.

\section{Differential \texorpdfstring{$\Zc$}{Z + c} cross section and \texorpdfstring{$\ZcZb$}{(Z + c)/(Z + b)} cross section ratio~\label{sec:xsec_diff}}

The $\Zc$ production cross section and the $\ZcZb$ cross section ratio are measured differentially as a function of the
transverse momentum of the $\PZ$ boson, $\ptZ$, and of the transverse momentum of the $\HF$ jet with the sample selected in the semileptonic mode
described in Section~\ref{sec:Zsel}.
The transverse momentum of the $\PZ$ boson is reconstructed from the momenta of the two selected leptons.
The sample is divided into three different subsamples according to the value of the variable of interest, $\ptZ$ or $\pt^{\jet}$,
and the fit procedure is performed independently for each of them and for each $\PZ$ boson decay mode.
The number and size of the bins is chosen such that the corrected secondary-vertex mass distribution for each bin is sufficiently populated to perform the signal extraction fit.

Potential effects of event migration between neighbouring bins and inside/outside the acceptance due to the detector resolution are studied using simulated samples.
A detector response matrix is built with those events fulfilling the selection criteria both with generated and reconstructed variables. The element $(i, j)$ in the matrix determines the probability that an event with generated $\ptZ$ ($\pt^{\jet}$) in bin $i$ ends up reconstructed in bin $j$ of the distribution.

Migration effects in $\ptZ$ are found to be negligible and no correction is applied.
An uncertainty of 1\%, which corresponds to the difference between the cross sections with and without corrections, is included in the systematic uncertainties.

Some migration of events between neighbouring bins in $\pt^{\jet}$ is expected because of the energy resolution, mainly between the first and second bins ($< 30\%$),
while migrations between the second and third bins are less than 10\%.
Migration effects are expected to be the same in the two $\cPZ$ boson decay modes.
The response matrix is used to unfold the fitted signal yields to actual signal yields at particle level.
Events with a generated $\pt^{\jet}$ outside the fiducial region and reconstructed inside it because of resolution effects are subtracted prior to the unfolding procedure.
Corrections are made for acceptance losses at the border of the kinematical region because of the detector resolution and reconstruction inefficiencies.
The unfolding is performed with an analytical inversion of the matrix defining the event migrations.
Statistical and systematic uncertainties are propagated through the unfolding procedure.

\begin{figure*}[!htb]
\begin{center}
\includegraphics[width=\cmsDoubleFigWidth]{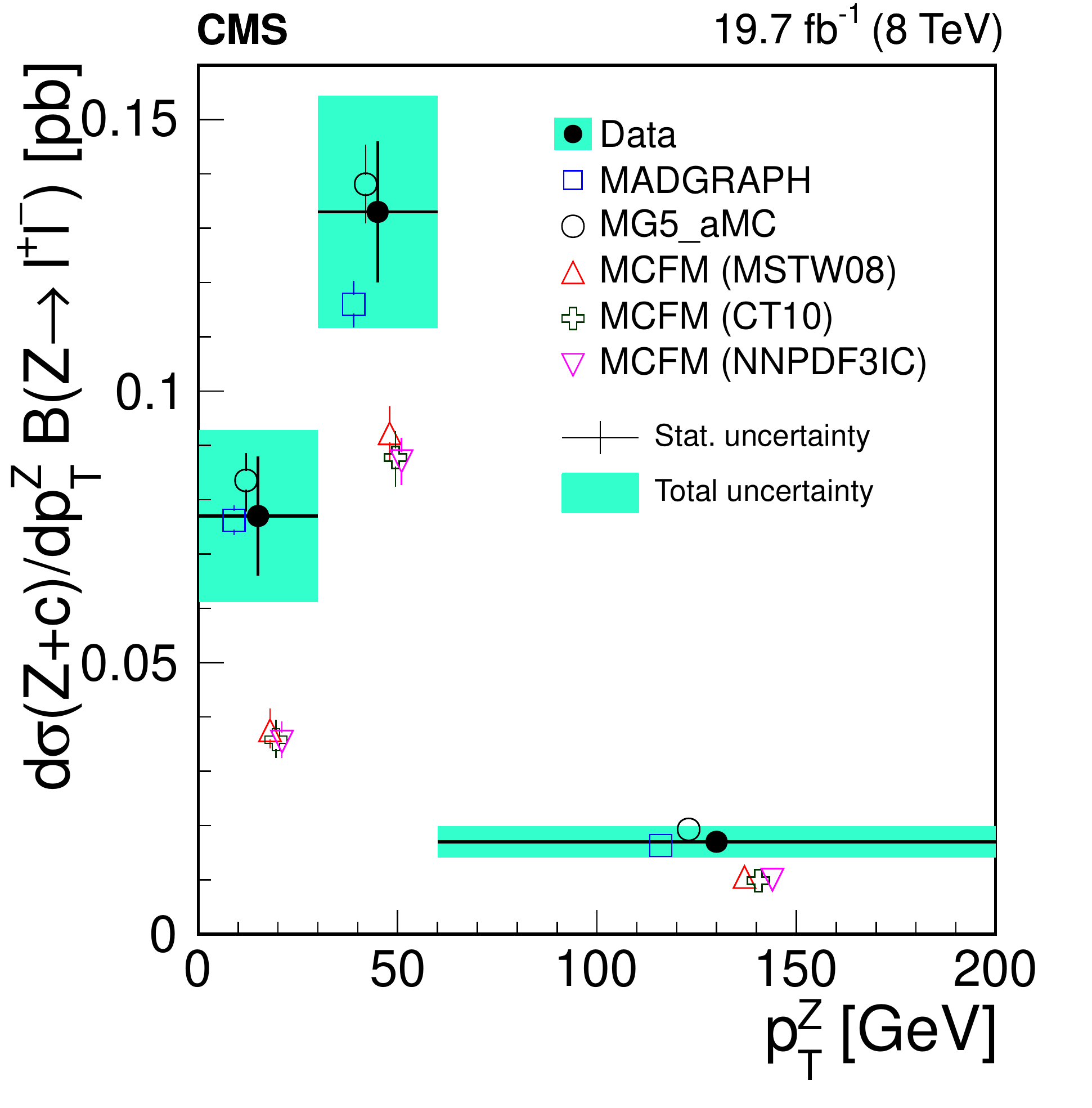}
\includegraphics[width=\cmsDoubleFigWidth]{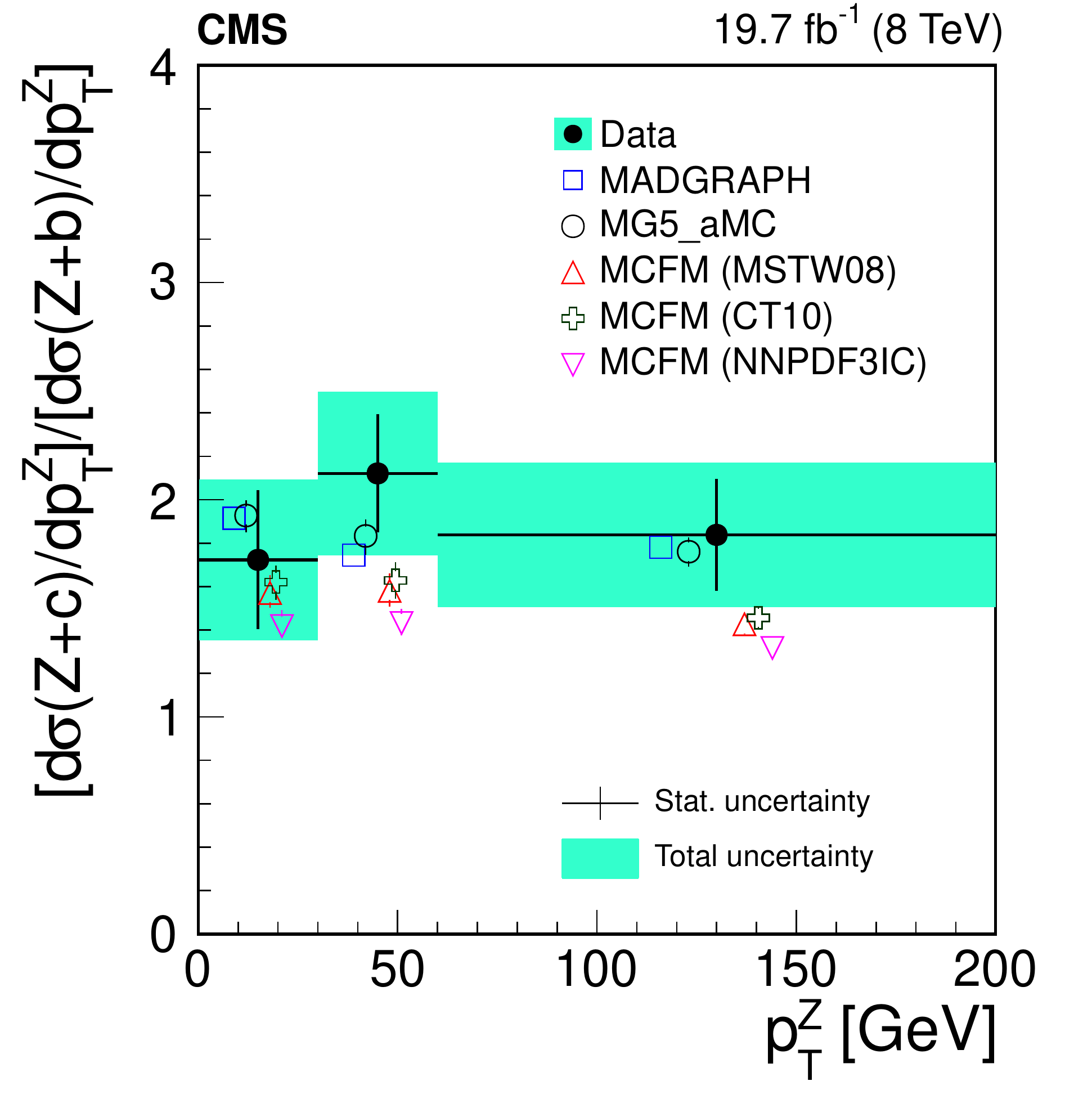} \\
\includegraphics[width=\cmsDoubleFigWidth]{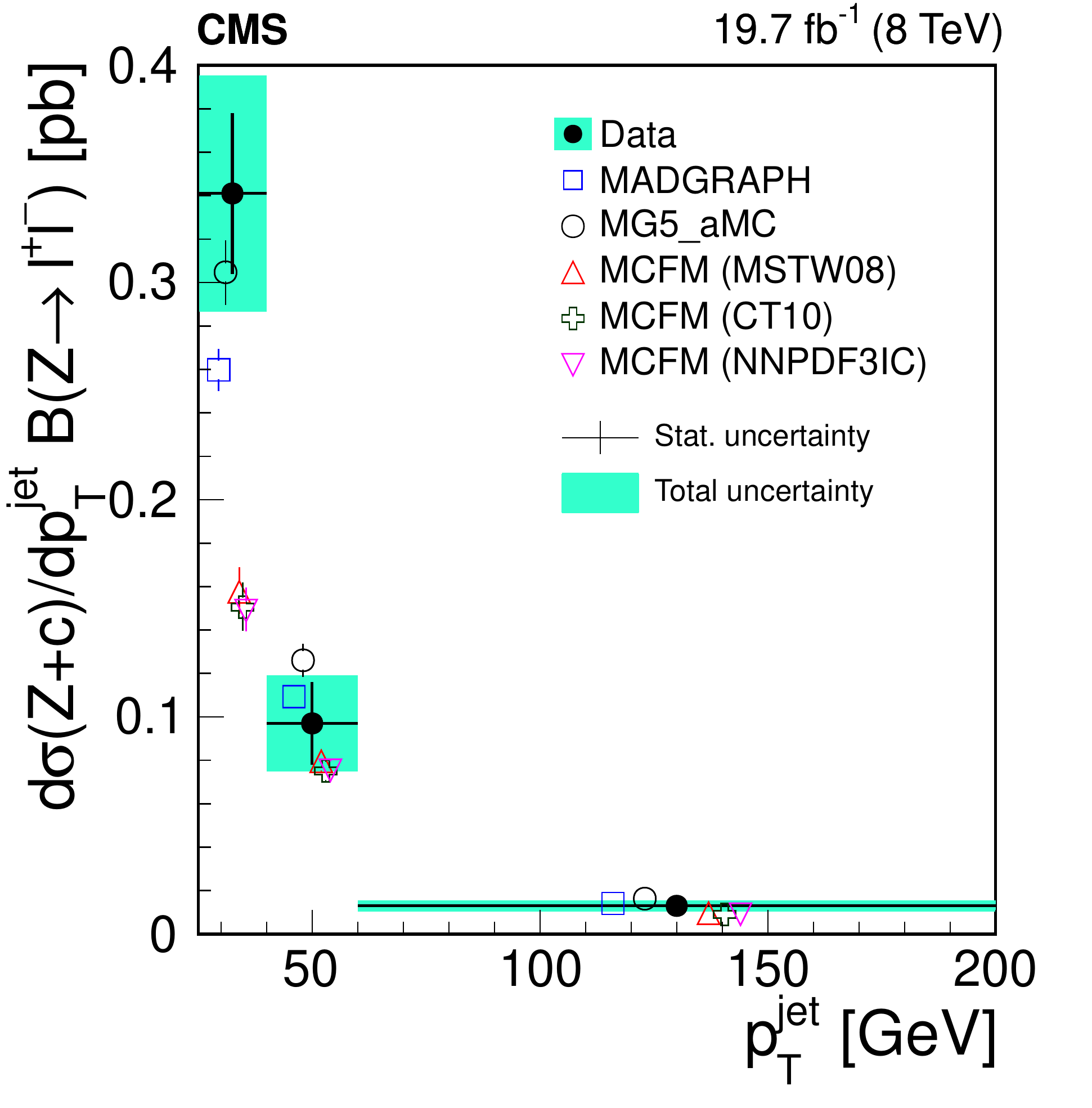}
\includegraphics[width=\cmsDoubleFigWidth]{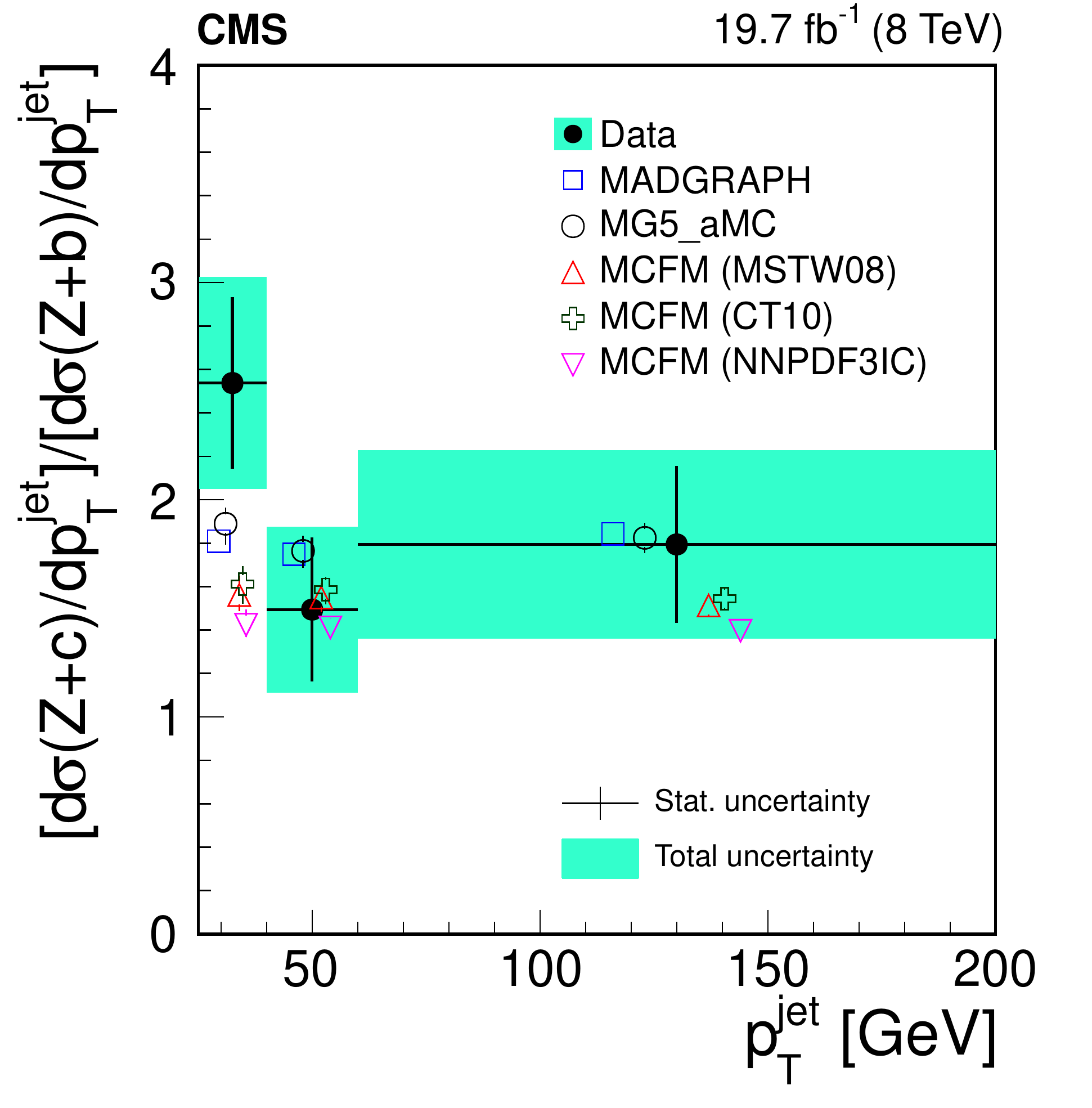}
\caption{Differential $\Zc$ cross section and $\ZcZb$ cross section ratio
as a function of the transverse momentum of the $\PZ$ boson (top) and the transverse momentum of the jet (bottom).
The combination of the results in the dielectron and dimuon channels is presented. The $\Zc$ differential cross section is shown on the left and
the $\ZcZb$ cross section ratio is on the right.
Statistical uncertainties in the data are shown as crosses. The solid rectangles indicate the total (statistical and systematic) experimental uncertainty.
Statistical and systematic uncertainties in the theoretical predictions are shown added in quadrature.
Symbols showing the theoretical expectations are slightly displaced from the bin centre in the horizontal axis for better visibility of the predictions.}
\label{fig:binnedjet_cs}
\end{center}
\end{figure*}

Tables~\ref{table:binned_Cross_sections} and~\ref{table:binnedjet_Cross_sections} summarize
the fitted $\Zc$ and $\Zb$ signal yields, the $\Zc$ cross section, and the $\ZcZb$ cross section ratio in the three $\ptZ$ and $\pt^{\jet}$ bins
and in the two $\PZ$ boson decay channels. The differential cross section and cross section ratio measured in the two $\PZ$ boson decay channels are consistent and are combined to obtain the final results, taking into account the statistical and systematic uncertainties in the two channels and the correlations among them.
The combined cross section and cross section ratio are presented in Table~\ref{table:combination_diff}.
They are also shown graphically in Fig.~\ref{fig:binnedjet_cs} in bins of $\ptZ$ (top) and $\pt^{\jet}$ (bottom).

\begin{table*}[tbp]
\begin{center}
\topcaption{Differential cross section $\SZcdiffpTZl\,\mathcal{B}$ and cross section ratio $(\SZcdiffpTZl)/(\SZbdiffpTZl)$ in the semileptonic mode and in the two Z boson decay channels.
The $N^{\rm signal}_{\Zc}$ and $N^{\rm signal}_{\Zb}$ are the yields of $\Zc$ and $\Zb$ events, respectively, extracted from the fit. All uncertainties quoted in the table are statistical, except for those of the measured cross sections and cross section ratios, where the first uncertainty is statistical and the second is the estimated systematic uncertainty from the sources discussed in the text.}
\renewcommand{\arraystretch}{1.2}
\begin{tabular}{c|c|c|c|c}
Channel & $N^{\rm signal}_{\Zc}$ & $\SZcdiffpTZ\,\mathcal{B}$ [pb] & $N^{\rm signal}_{\Zb}$ & $\SZcdiffpTZ/\SZbdiffpTZ$ \\ [1ex]
\hline
\multicolumn {5}{c}{$0<\ptZ<30\GeV$} \T\B  \\ \hline
$\PZ \to {\Pe^+}{\Pe^-}$ & 212 $\pm$  44 & 0.067 $\pm$ 0.014 $\pm$ 0.010 & 578 $\pm$ 52 & 1.5 $\pm$ 0.4 $\pm$ 0.2\\
$\PZ \to {\PGm^+}{\PGm^-}$ & 380 $\pm$  61 & 0.102 $\pm$ 0.016 $\pm$ 0.017 & 693 $\pm$ 68 & 2.7 $\pm$ 0.6 $\pm$ 0.4\\
\hline
\multicolumn {5}{c}{$30<\ptZ<60\GeV$} \T\B  \\ \hline
$\PZ \to {\Pe^+}{\Pe^-}$ & 501 $\pm$  60 & 0.144 $\pm$ 0.017 $\pm$ 0.019 & 1035 $\pm$ 66 & 2.4 $\pm$ 0.4 $\pm$ 0.3\\
$\PZ \to {\PGm^+}{\PGm^-}$ & 586 $\pm$  92 & 0.123 $\pm$ 0.019 $\pm$ 0.017 & 1422 $\pm$ 87 & 1.9 $\pm$ 0.4 $\pm$ 0.3\\
\hline
\multicolumn {5}{c}{$60<\ptZ<200\GeV$} \T\B  \\ \hline
$\PZ \to {\Pe^+}{\Pe^-}$ & 363 $\pm$ 53 & 0.017 $\pm$ 0.002 $\pm$ 0.002 &  913 $\pm$ 67 & 1.7 $\pm$ 0.3 $\pm$ 0.2\\
$\PZ \to {\PGm^+}{\PGm^-}$ & 474 $\pm$ 73 & 0.017 $\pm$ 0.003 $\pm$ 0.002 & 1056 $\pm$ 81 & 2.0 $\pm$ 0.4 $\pm$ 0.3\\
\end{tabular}
\label{table:binned_Cross_sections}
\end{center}
\end{table*}

\begin{table*}[tbp]
\begin{center}
\topcaption{Differential cross section $\SZcdiffpTjetl\,\mathcal{B}$ and cross section ratio $(\SZcdiffpTjetl)/(\SZbdiffpTjetl)$ in the semileptonic mode and in the two Z boson decay channels.
The $N^{\rm signal}_{\Zc}$ and $N^{\rm signal}_{\Zb}$ are the yields of $\Zc$ and $\Zb$ events, respectively, extracted from the fit. All uncertainties quoted in the table are statistical, except for those of the measured cross sections and cross section ratios, where the first uncertainty is statistical and the second is the estimated systematic uncertainty from the sources discussed in the text.}
\renewcommand{\arraystretch}{1.2}
\begin{tabular}{c|c|c|c|c}
Channel & $N^{\rm signal}_{\Zc}$ & $\SZcdiffpTjet\,\mathcal{B}$ [pb] & $N^{\rm signal}_{\Zb}$ & $\SZcdiffpTjet/\SZbdiffpTjet$ \\ [1.5ex]
\hline
\multicolumn {5}{c}{$25<\pt^{\jet}<40\GeV$} \T\B  \\ \hline
$\PZ \to {\Pe^+}{\Pe^-}$ & $476 \pm  58$ & $0.342 \pm 0.048 \pm 0.041$ & $1022 \pm 67$ & $2.7 \pm 0.6 \pm 0.3$\\
$\PZ \to {\PGm^+}{\PGm^-}$ & $583 \pm  91$ & $0.337 \pm 0.059 \pm 0.055$ & $1393\pm  90$ & $2.4 \pm 0.6 \pm 0.4$\\
\hline
\multicolumn {5}{c}{$40<\pt^{\jet}<60\GeV$} \T\B  \\ \hline
$\PZ \to {\Pe^+}{\Pe^-}$ & $289 \pm  47$ & $0.090 \pm 0.027 \pm 0.018$ & $ 843 \pm 59$ & $1.3 \pm 0.4 \pm 0.2$\\
$\PZ \to {\PGm^+}{\PGm^-}$ & $456 \pm  66$ & $0.104 \pm 0.027 \pm 0.014$ & $1044 \pm 75$ & $1.9 \pm 0.5 \pm 0.3$\\
\hline
\multicolumn {5}{c}{$60<\pt^{\jet}<200\GeV$} \T\B  \\ \hline
$\PZ \to {\Pe^+}{\Pe^-}$ & $311 \pm  56$ & $0.012 \pm 0.003 \pm 0.008$ & $686 \pm 64 $ & $1.7 \pm 0.5 \pm 0.3$\\
$\PZ \to {\PGm^+}{\PGm^-}$ & $369 \pm  63$ & $0.013 \pm 0.003 \pm 0.007$ & $800 \pm 75 $ & $1.9 \pm 0.5 \pm 0.3$\\
\end{tabular}
\label{table:binnedjet_Cross_sections}
\end{center}
\end{table*}

\begin{table*}[tbp]
\begin{center}
\topcaption{Differential $\Zc$ cross section and $\ZcZb$ cross section ratio. The first column presents the $\pt$ range for each bin.
Column 2 presents the cross section and column 3 the ratio.
The differential measurements as a function of the transverse momentum of the Z boson (jet with heavy flavour content) are given in the upper (lower) part of the table.
The first uncertainty is statistical and the second is the systematic uncertainty arising from the sources discussed in the text.
}
\renewcommand{\arraystretch}{1.2}
\begin{tabular}{c|c|c}
$[\ptZ{}_{\rm min}, \ptZ{}_\text{max}]$ \T & \multirow{2}{*}{$\SZcdiffpTZ\, \mathcal{B}$ [pb]} & \multirow{2}{*}{$\SZcdiffpTZ/\SZbdiffpTZ$} \\
$[\mathrm{GeV}]$ & & \\
\hline
$\x[0, 30]$   & $0.077 \pm 0.011 \pm 0.011$ & $1.7 \pm 0.3 \pm 0.2$ \\
$[30, 60]$  & $0.133 \pm 0.013 \pm 0.017$ & $2.1 \pm 0.3 \pm 0.3$ \\
$\x[60, 200]$ & $0.017 \pm 0.002 \pm 0.002$ & $1.8 \pm 0.3 \pm 0.2$ \\
\hline
$[\pt^{\jet}{}_{\rm min}, \pt^{\jet}{}_\text{max}]$ \T& \multirow{2}{*}{$\SZcdiffpTjet\, \mathcal{B}$ [pb]} & \multirow{2}{*}{$\SZcdiffpTjet/\SZbdiffpTjet$} \\
$[\mathrm{GeV}]$ & & \\
\hline
$[25, 40]$  & $0.341 \pm 0.037 \pm 0.040$ & $2.5 \pm 0.4 \pm 0.3$ \\
$[40, 60]$  & $0.097 \pm 0.019 \pm 0.012$ & $1.5 \pm 0.3 \pm 0.2$ \\
$\x[60, 200]$ & $0.013 \pm 0.002 \pm 0.002$ & $1.8 \pm 0.4 \pm 0.2$ \\
\end{tabular}
\label{table:combination_diff}
\end{center}
\end{table*}

Theoretical predictions for the differential cross section and cross section ratio are also obtained with the two MC generator programs and with \MCFM.
They are shown in Fig.~\ref{fig:binnedjet_cs} for comparison with the measured values.
The uncertainties in the \MADGRAPH predictions include the statistical and PDF uncertainties.
Scale variations are additionally included in the uncertainties of \aMCatNLO and \MCFM.
Predictions from \aMCatNLO are higher than the predictions from \MADGRAPH in the three bins of the $\Zc$ differential distributions.
A higher $\ZcZb$ cross section ratio is predicted up to $60 \GeV$, although consistent within uncertainties.
The predictions from \MADGRAPH and \aMCatNLO successfully reproduce the measurements.
The level of agreement is similar in terms of the $\Zc$ cross section and the $\ZcZb$ cross section ratio.

A similar ordering appears in the differential cross sections and the inclusive cross sections for theoretical predictions
calculated with \MCFM and the various PDF sets.
The highest $\Zc$ cross section is predicted using the MSTW08 PDF set,
the largest differential $\ZcZb$ cross section ratio in the two variables is obtained with the CT10 PDF set.
All \MCFM predictions are lower than the differential cross section measurements as a function of $\ptZ$.
This discrepancy is most pronounced in the first bin in $\pt^{\jet}$.
Differences between predictions and data are reduced in the $\ZcZb$ cross section ratio comparison.

The fitted charm PDF in NNPDF3IC~\cite{Ball:2016neh} set is consistent with having an intrinsic component.
The fitted fraction of the proton momentum that the charm quark component carries is $(0.7 \pm 0.3)\%$ if EMC data~\cite{Aubert:1982tt} is included in the fit
and $(1.6 \pm 1.2)\%$ without it.
After subtraction of the perturbative component, the momentum fraction of the proton carried by the IC component is $(0.5 \pm 0.3)\%$ if EMC data is
included in the fit, or $(1.4 \pm 1.2)\%$ if not.
Upper limits from the CTEQ-TEA Collaboration are also available~\cite{Dulat:2013hea, Hou:2015emq}.
Quoted limits on the proton momentum fraction carried by the IC component vary between 1.5\% and  2.5\% at 90\% confidence level
depending on the parameterization used.

If the proton momentum fraction taken by the charm quark component (intrinsic + perturbative) is of order ${\approx}2\%$,
an increase in the production of $\Zc$ events with a $\ptZ{\approx}100\GeV$ of at least 20--25\% would be expected~\cite{Ball:2016neh}.
Should it be smaller than 1\%, the cross section increase would be limited in the $\ptZ$ region around 100--200\GeV and only become visible at significantly
higher $\ptZ$ (${\gtrsim}500\GeV$). The measured cross section in the $\ptZ$ bin [60, 200]\GeV is in agreement with predictions from \MADGRAPH
and \aMCatNLO using a perturbative charm quark PDF. This measurement is in agreement with no increase in the production rate
or with a very modest one, as expected from current upper limits on the IC component. No increase in the production rate in the highest $\pt^{\jet}$ bin is observed, either.

\section{Summary}
\label{sec:summary}

The associated production of a $\PZ$ boson with at least one charm quark jet
in proton-proton collisions at a centre-of-mass energy of 8 \TeV was studied with a data sample corresponding to an integrated luminosity of $19.7 \pm 0.5\fbinv$.
It was compared to the production of a $\PZ$ boson with at least one $\PQb$ quark jet.
Selection of event candidates relies on the identification of semileptonic decays of $\PQc$ or $\bhadrons$ with a muon in the final state and through the reconstruction
of exclusive decay channels of $\Dpm$ and $\Dstar$ mesons. The $\PZ$ boson is identified through its decay into an $\Pe^+\Pe^-$ or $\PGm^+\PGm^-$ pair.

The cross section for the production of a $\PZ$ boson associated with at least one $\PQc$ quark jet is measured.
The measurement is performed in the kinematic region with two leptons with transverse momentum $\pt^{\ell}>20\GeV$, pseudorapidity $\abs{\eta^{\ell}} < 2.1$,
dilepton invariant mass $71 < m_{\ell\ell} < 111\GeV$ and a jet with $\pt^{\jet}>25\GeV$, $\abs{\eta^{\jet}} <2.5$, separated from the leptons
of the $\PZ$ boson candidate by a distance $\Delta R ({\text{jet}},\ell) > 0.5$.

The $\Zc$ production cross sections measured in all the analysis categories are fully consistent, and the combined value is
$\sigma(\ppZc) \mathcal{B}(\PZ \to \ell^+\ell^-) = 8.8 \pm 0.5 \stat \pm 0.6 \syst \unit{pb}$.
This is the first measurement at the LHC of $\Zc$ production in the central pseudorapidity region.

The cross section ratio for the production of a $\PZ$ boson and at least one $\PQc$ and at least one $\PQb$ quark jet is
measured in the same kinematic region and is $\sigma(\ppZc)/\sigma(\ppZb) = 2.0 \pm 0.2 \stat \pm 0.2 \syst$.

The size of the sample selected in the semileptonic channel allows for the first differential measurements of the $\Zc$ cross section at the LHC.
The $\Zc$ cross section and $\ZcZb$ cross section ratio are measured as a function of the transverse momentum of the $\PZ$ boson and of the
heavy flavour jet.

The measurements are in agreement with the leading order predictions from \MADGRAPH and next-to-leading-order predictions from \MGvATNLO.
Predictions from the \MCFM program are lower than the measured $\Zc$ cross section and $\ZcZb$ cross section ratio, both inclusively and differentially.
This difference can be explained by the absence of parton shower development and nonperturbative effects in the \MCFM calculation.

Measurements in the highest $\ptZ$ ($\pt^{\jet}$) region analyzed, $60 < \ptZ (\pt^{\jet}) < 200 \GeV$, would be sensitive to
the existence of an intrinsic charm component inside the proton if this IC component were large enough to induce a significant enhancement in the
$\Zc$ production cross section.
However, our measurements of the $\Zc$ cross section and $\ZcZb$ cross section ratio are consistent with predictions using PDF sets with no
IC component.

\begin{acknowledgments}
We congratulate our colleagues in the CERN accelerator departments for the excellent performance of the LHC and thank the technical and administrative staffs at CERN and at other CMS institutes for their contributions to the success of the CMS effort. In addition, we gratefully acknowledge the computing centres and personnel of the Worldwide LHC Computing Grid for delivering so effectively the computing infrastructure essential to our analyses. Finally, we acknowledge the enduring support for the construction and operation of the LHC and the CMS detector provided by the following funding agencies: BMWFW and FWF (Austria); FNRS and FWO (Belgium); CNPq, CAPES, FAPERJ, and FAPESP (Brazil); MES (Bulgaria); CERN; CAS, MoST, and NSFC (China); COLCIENCIAS (Colombia); MSES and CSF (Croatia); RPF (Cyprus); SENESCYT (Ecuador); MoER, ERC IUT, and ERDF (Estonia); Academy of Finland, MEC, and HIP (Finland); CEA and CNRS/IN2P3 (France); BMBF, DFG, and HGF (Germany); GSRT (Greece); OTKA and NIH (Hungary); DAE and DST (India); IPM (Iran); SFI (Ireland); INFN (Italy); MSIP and NRF (Republic of Korea); LAS (Lithuania); MOE and UM (Malaysia); BUAP, CINVESTAV, CONACYT, LNS, SEP, and UASLP-FAI (Mexico); MBIE (New Zealand); PAEC (Pakistan); MSHE and NSC (Poland); FCT (Portugal); JINR (Dubna); MON, RosAtom, RAS, RFBR and RAEP (Russia); MESTD (Serbia); SEIDI, CPAN, PCTI and FEDER (Spain); Swiss Funding Agencies (Switzerland); MST (Taipei); ThEPCenter, IPST, STAR, and NSTDA (Thailand); TUBITAK and TAEK (Turkey); NASU and SFFR (Ukraine); STFC (United Kingdom); DOE and NSF (USA).

\hyphenation{Rachada-pisek} Individuals have received support from the Marie-Curie programme and the European Research Council and Horizon 2020 Grant, contract No. 675440 (European Union); the Leventis Foundation; the A. P. Sloan Foundation; the Alexander von Humboldt Foundation; the Belgian Federal Science Policy Office; the Fonds pour la Formation \`a la Recherche dans l'Industrie et dans l'Agriculture (FRIA-Belgium); the Agentschap voor Innovatie door Wetenschap en Technologie (IWT-Belgium); the Ministry of Education, Youth and Sports (MEYS) of the Czech Republic; the Council of Science and Industrial Research, India; the HOMING PLUS programme of the Foundation for Polish Science, cofinanced from European Union, Regional Development Fund, the Mobility Plus programme of the Ministry of Science and Higher Education, the National Science Center (Poland), contracts Harmonia 2014/14/M/ST2/00428, Opus 2014/13/B/ST2/02543, 2014/15/B/ST2/03998, and 2015/19/B/ST2/02861, Sonata-bis 2012/07/E/ST2/01406; the National Priorities Research Program by Qatar National Research Fund; the Programa Clar\'in-COFUND del Principado de Asturias; the Thalis and Aristeia programmes cofinanced by EU-ESF and the Greek NSRF; the Rachadapisek Sompot Fund for Postdoctoral Fellowship, Chulalongkorn University and the Chulalongkorn Academic into Its 2nd Century Project Advancement Project (Thailand); the Welch Foundation, contract C-1845; and the Weston Havens Foundation (USA). \end{acknowledgments}

\bibliography{auto_generated}
\cleardoublepage \appendix\section{The CMS Collaboration \label{app:collab}}\begin{sloppypar}\hyphenpenalty=5000\widowpenalty=500\clubpenalty=5000\textbf{Yerevan Physics Institute,  Yerevan,  Armenia}\\*[0pt]
A.M.~Sirunyan, A.~Tumasyan
\vskip\cmsinstskip
\textbf{Institut f\"{u}r Hochenergiephysik,  Wien,  Austria}\\*[0pt]
W.~Adam, F.~Ambrogi, E.~Asilar, T.~Bergauer, J.~Brandstetter, E.~Brondolin, M.~Dragicevic, J.~Er\"{o}, M.~Flechl, M.~Friedl, R.~Fr\"{u}hwirth\cmsAuthorMark{1}, V.M.~Ghete, J.~Grossmann, N.~H\"{o}rmann, J.~Hrubec, M.~Jeitler\cmsAuthorMark{1}, A.~K\"{o}nig, I.~Kr\"{a}tschmer, D.~Liko, T.~Madlener, T.~Matsushita, I.~Mikulec, E.~Pree, D.~Rabady, N.~Rad, H.~Rohringer, J.~Schieck\cmsAuthorMark{1}, M.~Spanring, D.~Spitzbart, J.~Strauss, W.~Waltenberger, J.~Wittmann, C.-E.~Wulz\cmsAuthorMark{1}, M.~Zarucki
% ...
\vskip\cmsinstskip
1:~~Also at Vienna University of Technology, Vienna, Austria\\
% ...

\end{sloppypar}
\end{document}